\documentclass[journal,twoside,web]{ieeecolor}
\usepackage{generic}
\usepackage{cite}

\usepackage{amsmath, amsthm, amssymb, amsfonts}
\usepackage{graphicx}
\usepackage{textcomp}
\usepackage[resetlabels,labeled]{multibib}
\newcites{A}{A References}

\usepackage{flushend}       
\usepackage{array}
\usepackage{mdwtab}
\usepackage{eqparbox}
\usepackage{url}
\usepackage[export]{adjustbox}
\usepackage{wrapfig}
\usepackage{lipsum}
\usepackage{graphicx,subfigure}
\usepackage{adjustbox}
\usepackage[normalem]{ulem}
\useunder{\uline}{\ul}{}
\usepackage{tablefootnote}

\usepackage{pdflscape}
\usepackage{afterpage}

\usepackage{graphicx}
\usepackage{booktabs,makecell}
\usepackage{pdflscape}
\usepackage{caption}
\usepackage{mathtools}
\usepackage{subfigure}
\usepackage{comment}
\usepackage{amsmath}
\usepackage{longtable}
\usepackage[ruled,linesnumbered]{algorithm2e}
\usepackage{siunitx}
\usepackage{placeins}

\usepackage{multirow}
\usepackage{xr-hyper}
\usepackage{hyperref}

\theoremstyle{definition}

\graphicspath{ {./images/} }

\usepackage{scalerel}
\usepackage{tikz}
\usetikzlibrary{svg.path}
\definecolor{orcidlogocol}{HTML}{A6CE39}
\tikzset{
  orcidlogo/.pic={
    \fill[orcidlogocol] svg{M256,128c0,70.7-57.3,128-128,128C57.3,256,0,198.7,0,128C0,57.3,57.3,0,128,0C198.7,0,256,57.3,256,128z};
    \fill[white] svg{M86.3,186.2H70.9V79.1h15.4v48.4V186.2z}
                 svg{M108.9,79.1h41.6c39.6,0,57,28.3,57,53.6c0,27.5-21.5,53.6-56.8,53.6h-41.8V79.1z M124.3,172.4h24.5c34.9,0,42.9-26.5,42.9-39.7c0-21.5-13.7-39.7-43.7-39.7h-23.7V172.4z}
                 svg{M88.7,56.8c0,5.5-4.5,10.1-10.1,10.1c-5.6,0-10.1-4.6-10.1-10.1c0-5.6,4.5-10.1,10.1-10.1C84.2,46.7,88.7,51.3,88.7,56.8z};
  }
}

\newcommand\orcidicon[1]{\href{https://orcid.org/#1}{\mbox{\scalerel*{
\begin{tikzpicture}[yscale=-1,transform shape]
\pic{orcidlogo};
\end{tikzpicture}
}{|}}}}

\def\BibTeX{{\rm B\kern-.05em{\sc i\kern-.025em b}\kern-.08em
    T\kern-.1667em\lower.7ex\hbox{E}\kern-.125emX}}
\begin{document}
\title{Biceph-Net: A robust and lightweight framework for the diagnosis of Alzheimer's disease using 2D-MRI scans and deep similarity learning}
\author{A.H. Rashid, A. Gupta, J. Gupta, M. Tanveer$^*$ \orcidicon{0000-0002-5727-3697}  
   \thanks{
   \noindent $^*$ Corresponding Author\\
   M. Tanveer, A. H. Rashid, A. Gupta and J. Gupta are with the Department of Mathematics, Indian Institute of Technology Indore, Simrol, Indore, 453552, India (e-mail (M. Tanveer): mtanveer@iiti.ac.in, e-mail (A.H. Rashid): ashrafrashid@iiti.ac.in, e-mail (A. Gupta): aditya.fhd@gmail.com, e-mail (J. Gupta): gupta.jhalak00@gmail.com}
 }  

\maketitle

\begin{abstract}
Alzheimer's Disease (AD) is a neurodegenerative disease that is one of the significant causes of death in the elderly population. Many deep learning techniques have been proposed to diagnose AD using Magnetic Resonance Imaging (MRI) scans. Predicting AD using 2D slices extracted from 3D MRI scans is challenging as the inter-slice information gets lost. To this end, we propose a novel and lightweight framework termed `Biceph-Net' for AD diagnosis using 2D MRI scans that model both the intra-slice and inter-slice information. `Biceph-Net' has been experimentally shown to perform similar to other Spatio-temporal neural networks while being computationally more efficient. Biceph-Net is also superior in performance compared to vanilla 2D convolutional neural networks (CNN) for AD diagnosis using 2D MRI slices. Biceph-Net also has an inbuilt neighbourhood-based model interpretation feature that can be exploited to understand the classification decision taken by the network. Biceph-Net experimentally achieves a test accuracy of 100\% in the classification of Cognitively Normal (CN) vs AD,  98.16\% for Mild Cognitive Impairment (MCI) vs AD, and 97.80\% for CN vs MCI vs AD.
\end{abstract}

\vspace{4mm}
\section{Introduction}
\label{introduction}
\IEEEPARstart{A}lzheimer's Disease (AD) is a neurodegenerative disease that is one of the significant causes of death in the elderly population. Diagnosing AD is not trivial, even for a highly experienced medical practitioner, as it requires a precise examination of patient data. Typically, Cognitively Normal (CN) subjects progress to a pre-clinical phase with underlying biomarker abnormalities, then a prodromal state of Mild Cognitive Impairment (MCI) stage that affects memory, judgment, thinking and language before progressing to the full-fledged AD dementia stage. AD diagnosis through machine learning techniques has given tremendous results and is trending. The availability of a large amount of openly available data from various websites like Alzheimer's Disease Neuroimaging Initiative (ADNI), Australian Imaging, Biomarkers \& Lifestyle Flagship Study of Ageing (AIBL), and Open Access Series of Imaging Studies (OASIS) has further encouraged researchers to build more robust machine learning models for diagnosing AD. Classical machine learning techniques require hand-crafted features, which may not give an optimal result. Contrary to this, deep learning techniques automatically learn to extract features from the data and have performed marvellously for AD diagnosis from multi-modal data \cite{tanveer2020machine}.

\begin{figure*}
\centering     
\subfigure[]{\label{fig:intro_th_1}\includegraphics[width=5cm]{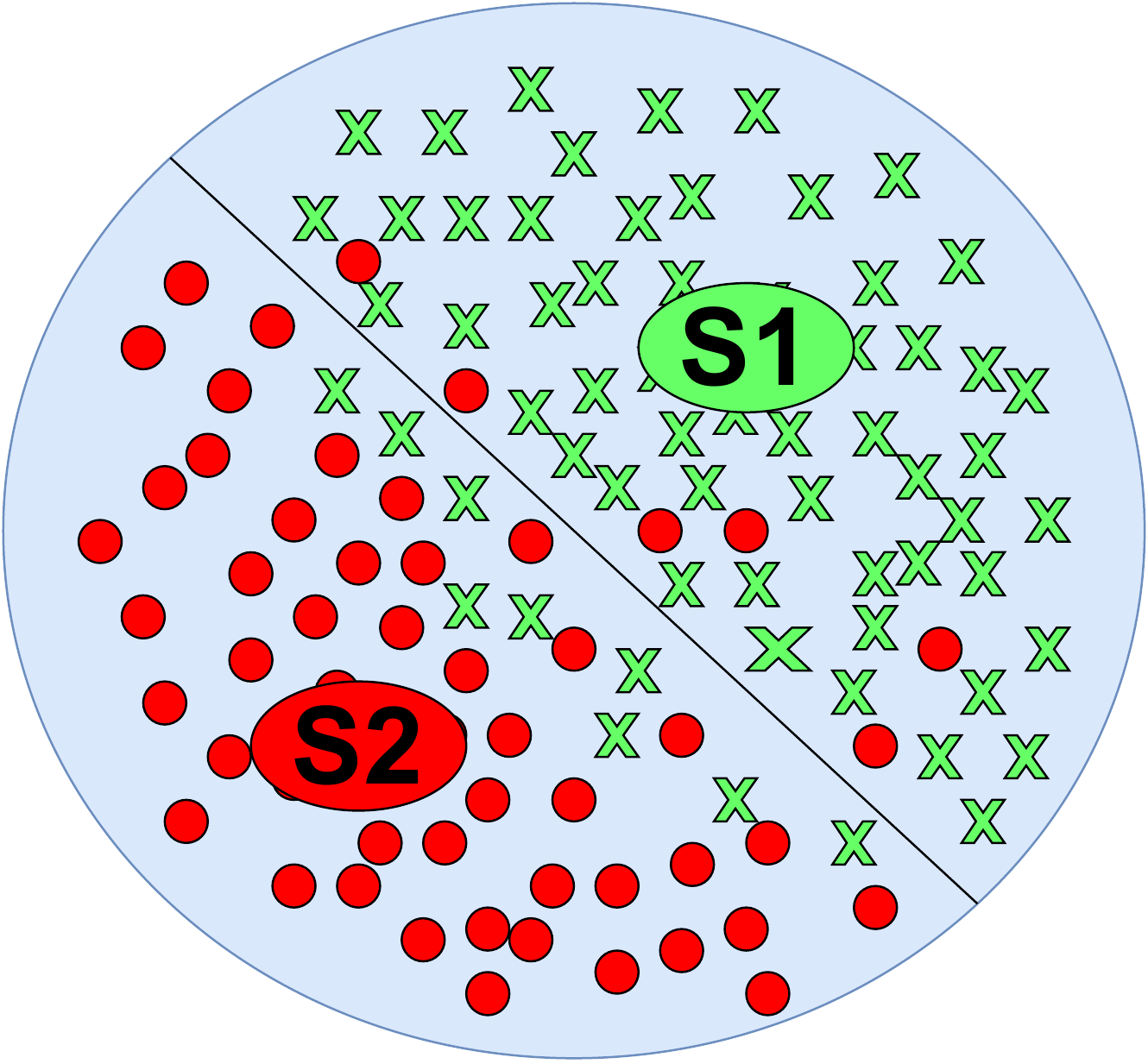}}
\subfigure[]{\label{fig:intro_th_2}\includegraphics[width=5cm]{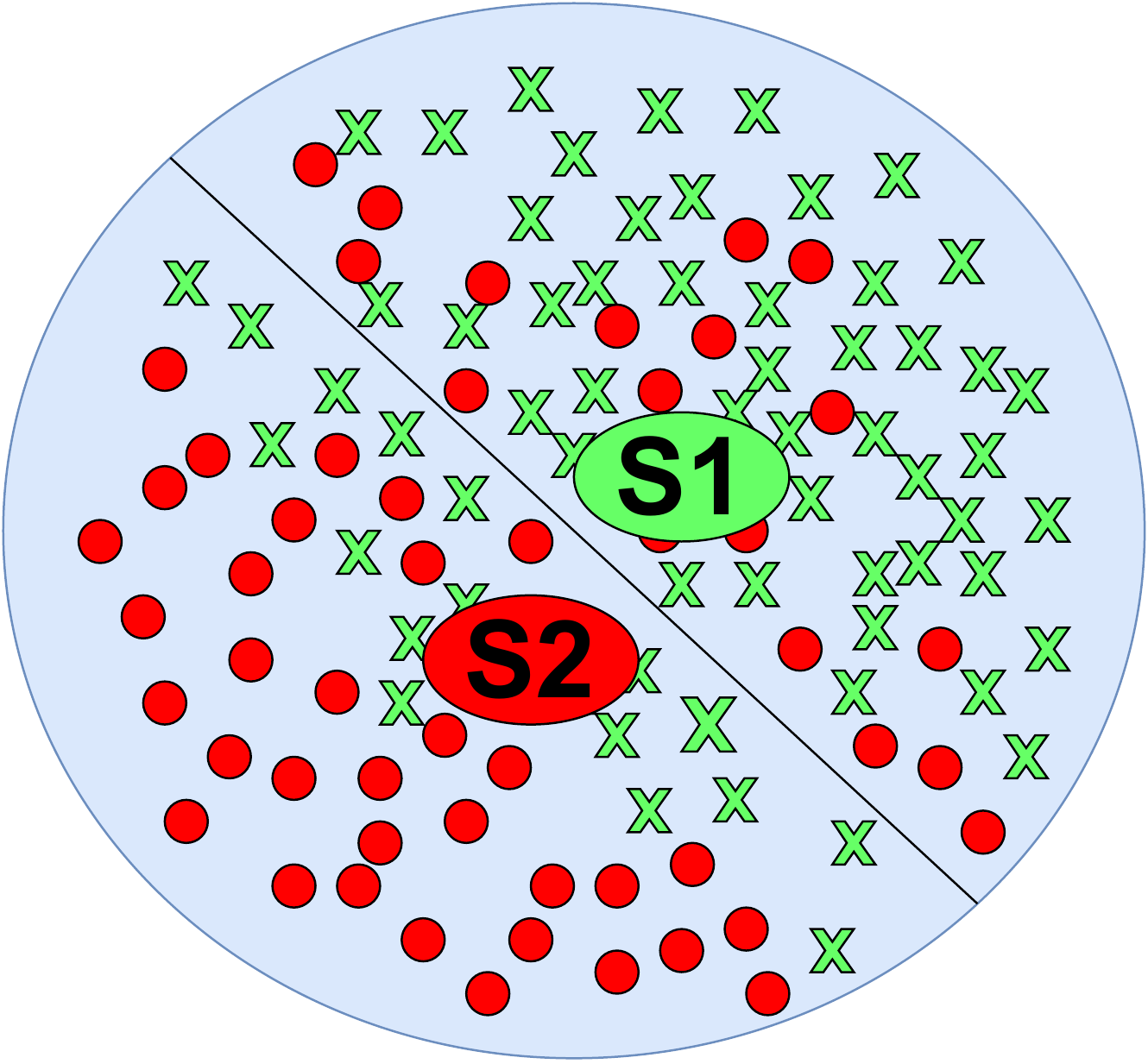}}
\subfigure[]{\label{fig:intro_th_3}\includegraphics[width=5cm]{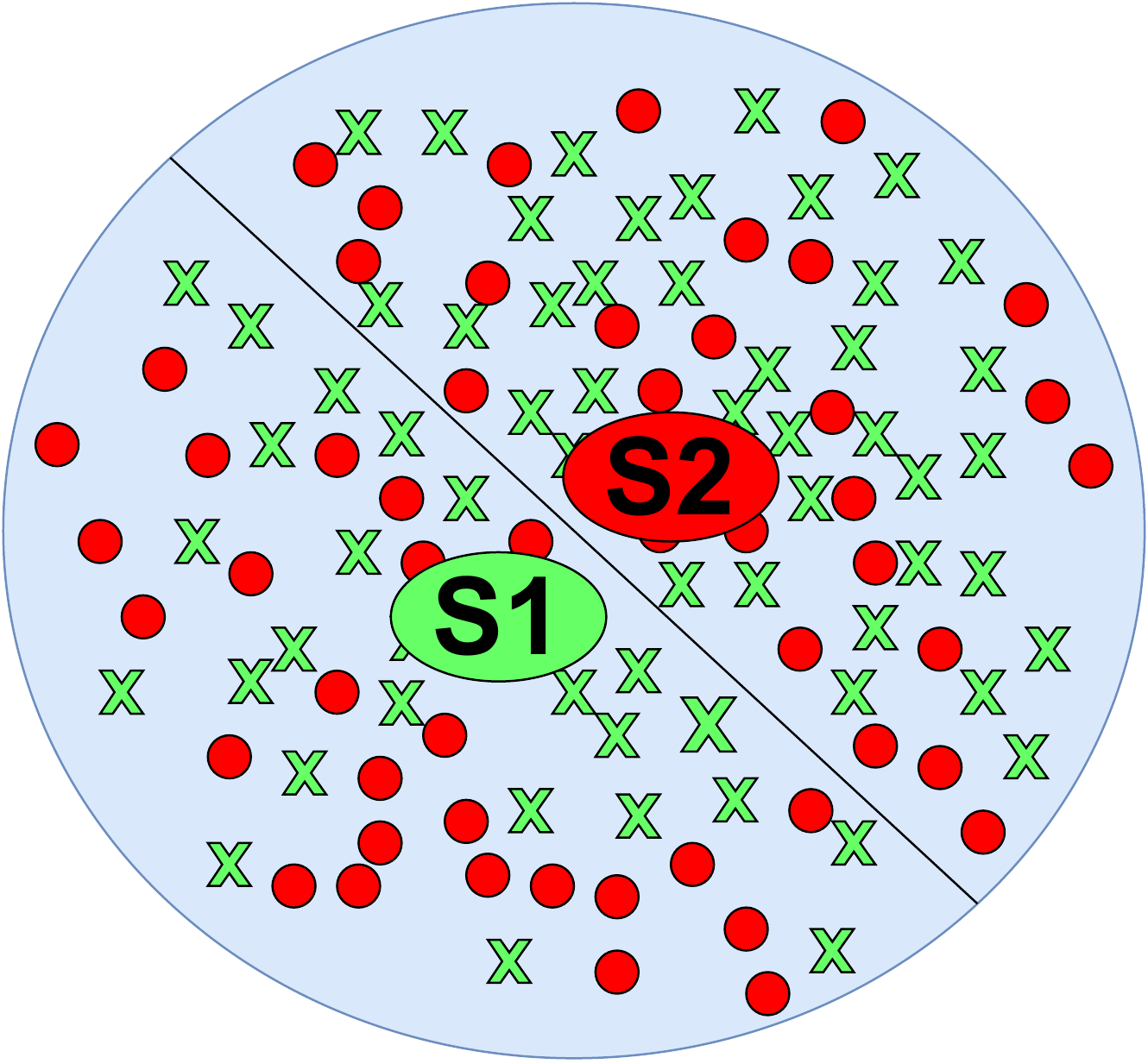}}
\caption{The process of classification of two subjects with multiple 2D-MRI scans. Figure (a) represents a correct subject-wise classification. Figure (b) also represents a correct subject-wise classification with higher slice-wise misclassifications. Figure (c) represents a subject-wise misclassification.}
\label{spm:stats_WM1}
\end{figure*}

With the growth in popularity of deep learning techniques in the classification of images, several different models were proposed to detect Alzheimer's disease using brain Magnetic Resonance Imaging (MRI). There are three main categories of neural networks that researchers have been using for the diagnosis of AD and its sub-types \cite{wen2020convolutional},  namely, (a) 3D Convolutional neural networks (3D-CNNs), (b) 2D Convolutional neural networks (2D-CNNs), and (c) Spatio-temporal convolutional neural networks (CNN-RNNs). In (a), the CNNs take 3D-MRI scans as input, whereas in (b) and (c), the CNNs take 2D-MRI slices extracted from a 3D-MRI scan as input. In this work, we mainly focus on diagnosing AD through the use of 2D-MRI scans. We utilise different types of networks that come under (b) and (c) to achieve the aforementioned objective.

The usage of 2D-MRI scans can prove beneficial in terms of the computation cost as 3D-CNNs are more computationally intensive than their 2D counterparts. However, since depth information is not available in 2D-MRI scans, 3D-CNNs tend to outperform 2D-CNNs when appropriately trained with sufficient data. The lack of depth information renders 2D-CNNs highly ineffective in the neuroimaging domain, where most of the data may be available in the form of 3D scans. To bridge this information gap, researchers use a combination of 2D-CNN and Recurrent Neural Networks (RNNs) that can learn the information present within individual 2D-MRI scans (intra-slice information), and among a sequence of 2D scans (inter-slice information) \cite{liu2018classification, el2020alzheimer, dua2020cnn}. The trade-off for utilising this extra information is the computational cost, as CNN-RNNs are more computationally intensive than vanilla CNNs \cite{hon2017towards}. 

In this work, we propose a robust and lightweight framework for training vanilla 2D CNNs to diagnose AD using 2D-MRI scans. We show that the proposed method is as good as CNN-RNNs in terms of classification accuracy while being more computationally efficient. We also show that the proposed method outperforms previously used 2D-CNNs for AD diagnosis. The proposed method also avoids the problem of leakage that was inherent in many of the previously used 2D-CNNs for the diagnosis of AD \cite{wen2020convolutional}.

In order to utilise 2D-CNNs for final classification, we extract 2D slices from a 3D-MRI scan. This renders the data into a hierarchical structure as we need to classify individual slices (slice-wise classification) and correctly classify each subject (subject-wise classification). With this type of structure, the classifier can afford to misclassify individual slices as long as the subject is classified correctly. However, a vanilla CNN guided by a cross-entropy loss cannot fully handle this task as it only looks at slice-wise classification. These networks may work well on the data in which the majority of slices can be easily classified, as shown in fig \ref{fig:intro_th_1}. The two classes are represented in green and red colours. S1 and S2 represent the mean of the green and red classes' samples, respectively.

As the similarity between individual slices of multiple subjects increases, the vanilla 2D-CNN tends to misclassify many individual slices to a level where entire subjects get misclassified, as shown in figures \ref{fig:intro_th_2} and \ref{fig:intro_th_3}. In \ref{fig:intro_th_2}, the slice-wise misclassifications are still within the threshold, which does not affect subject-wise classification. However, in \ref{fig:intro_th_3}, the slice-wise misclassifications go beyond the threshold and cause subject-wise misclassification. Therefore, for an effective classification using 2D-MRI slices, the classifier needs to:
\begin{enumerate}
    \item Perform slice-wise classification reasonably well. \label{imp1}
    \item Perform slice-wise misclassifications (if any) within a threshold such that the subject-wise classification is not impeded. \label{imp2}
\end{enumerate}

\begin{figure*}
\centering     

\subfigure[]{\label{fig:triplet_loss1}\includegraphics[width=5cm]{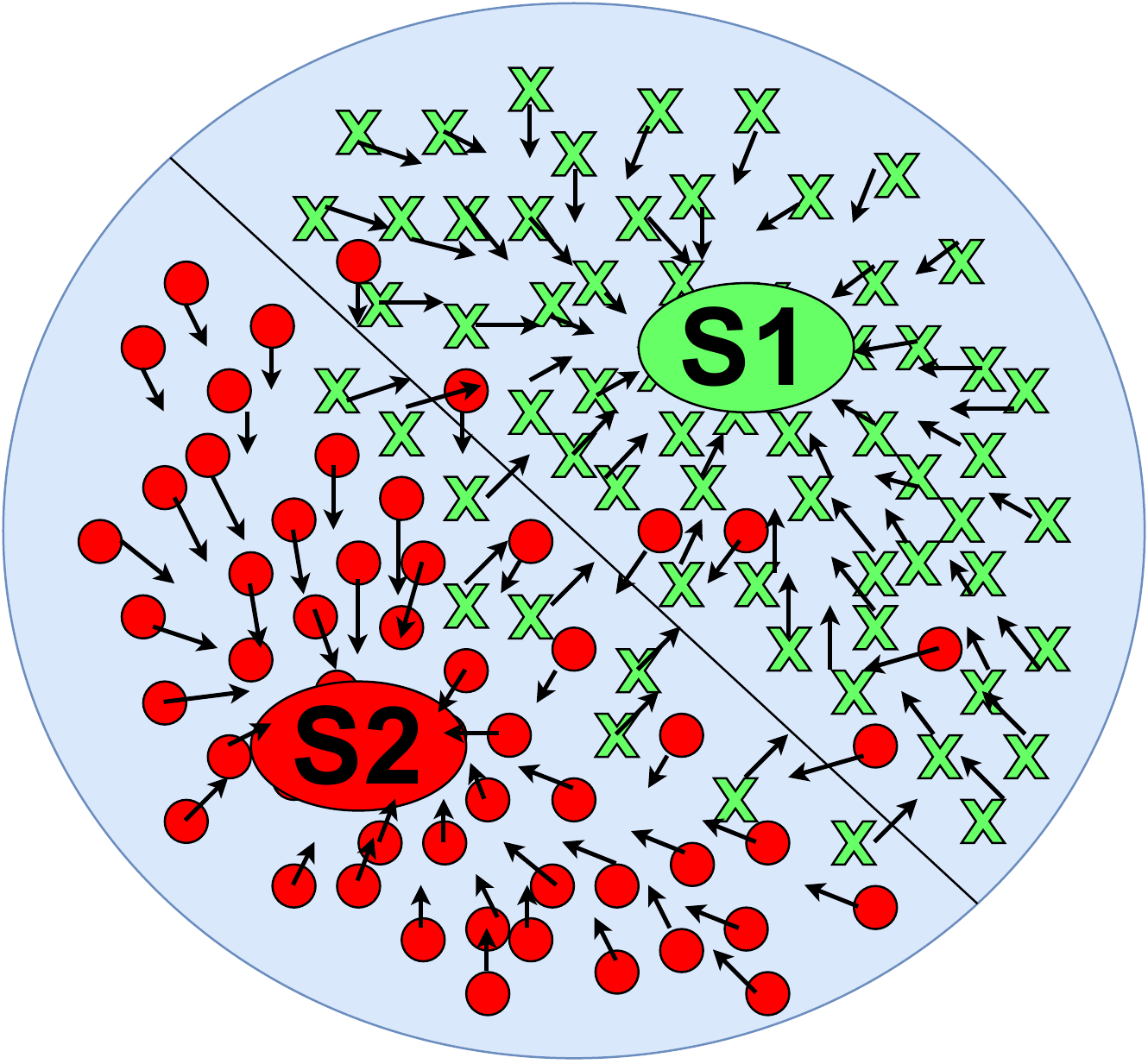}}
\subfigure[]{\label{fig:triplet_loss2}\includegraphics[width=5cm]{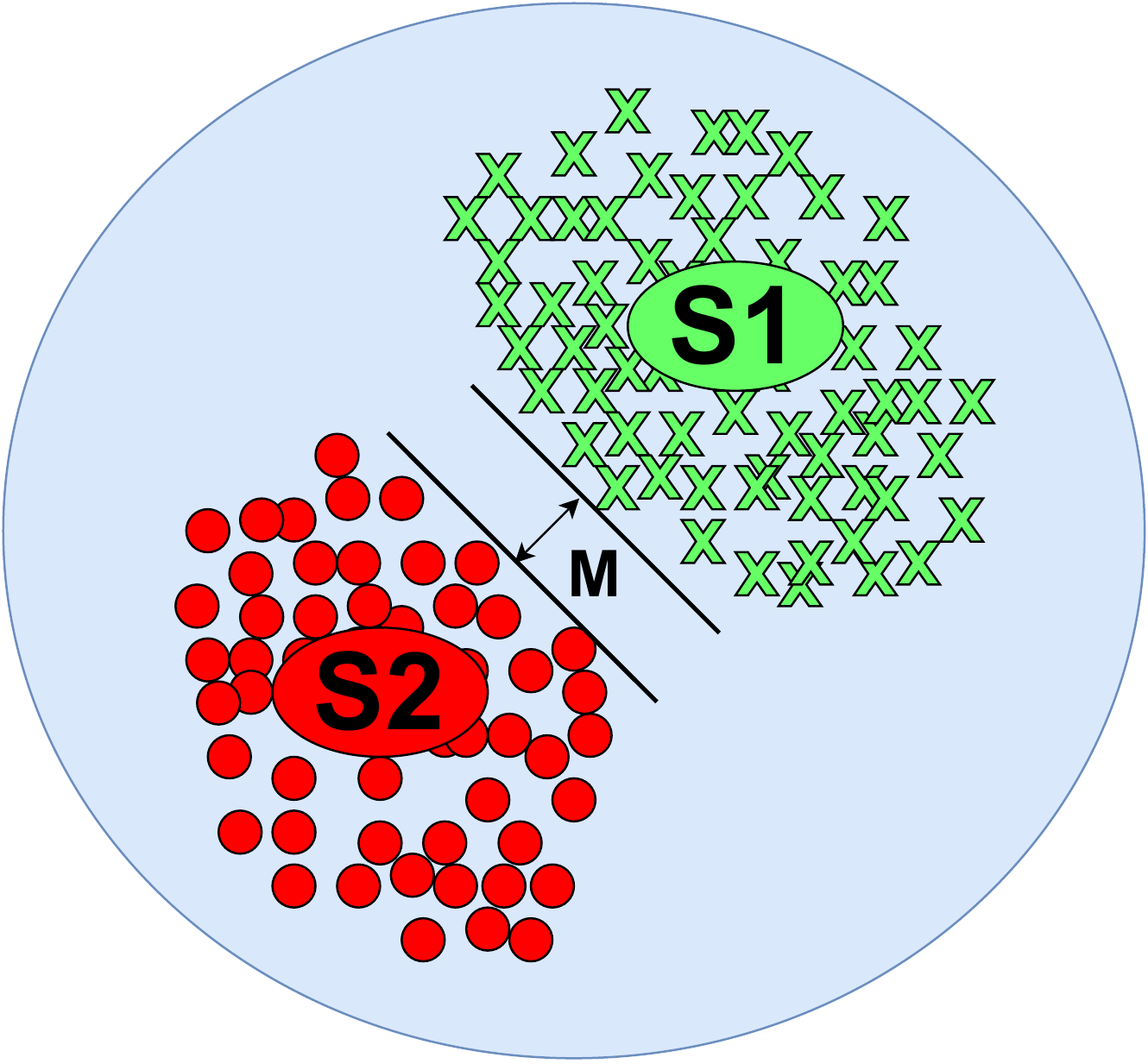}}
\subfigure[]{\label{fig:triplet_loss3}\includegraphics[width=5cm]{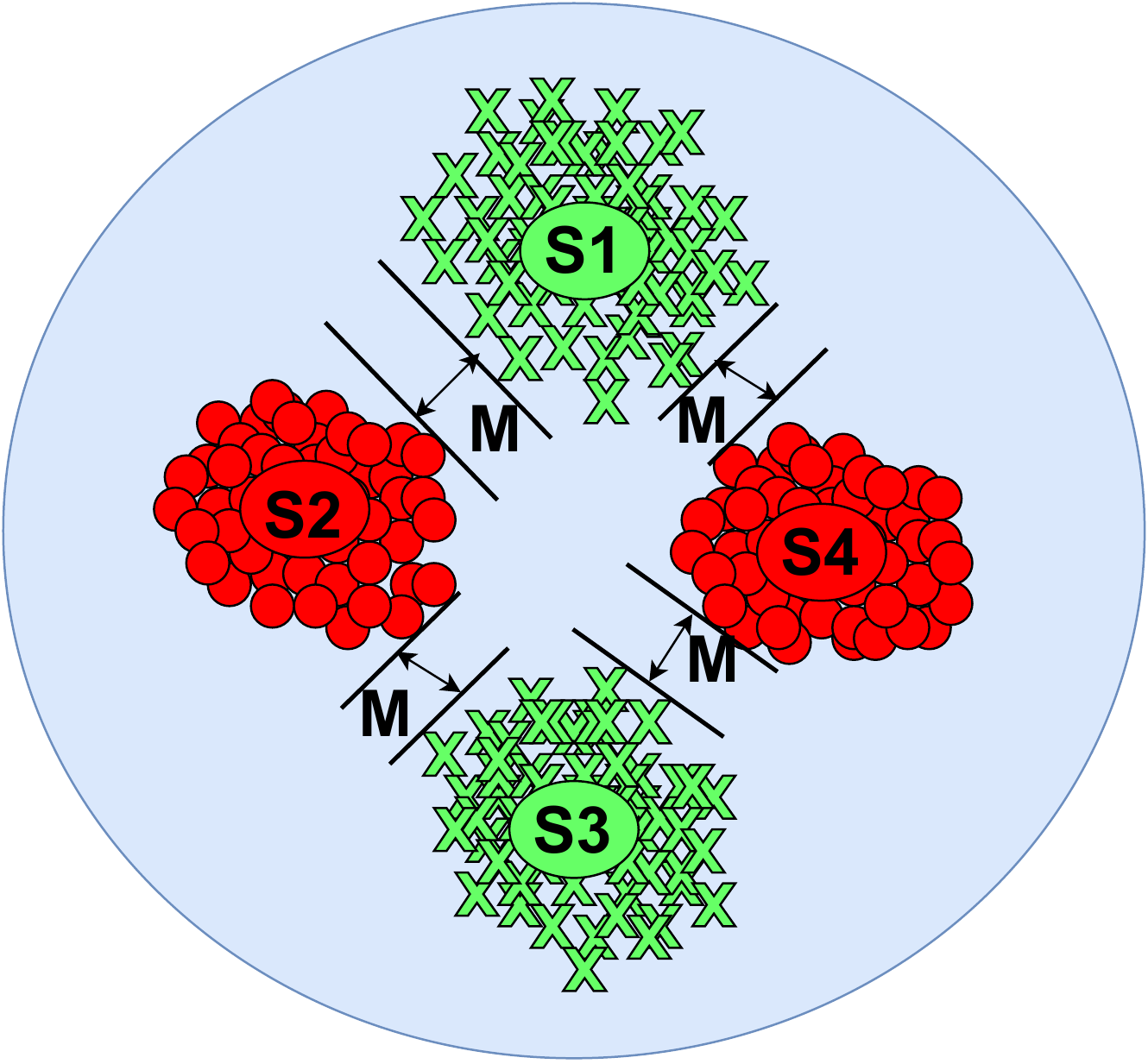}}
\caption{The effect of using Triplet loss on multiple 2D-MRI slices of two subjects. Figure (a) intuitively shows the slices of same subject getting an embedding close to each other. Figure (b) shows the embedding space after training is completed using Triplet loss. Figure (c) shows the embedding space with more than two subjects after training is completed using Triplet loss.}
\label{spm:stats_WM2}
\end{figure*}

\begin{figure*}
\centering

\subfigure[]{\label{fig:biceph_loss1}\includegraphics[width=5cm]{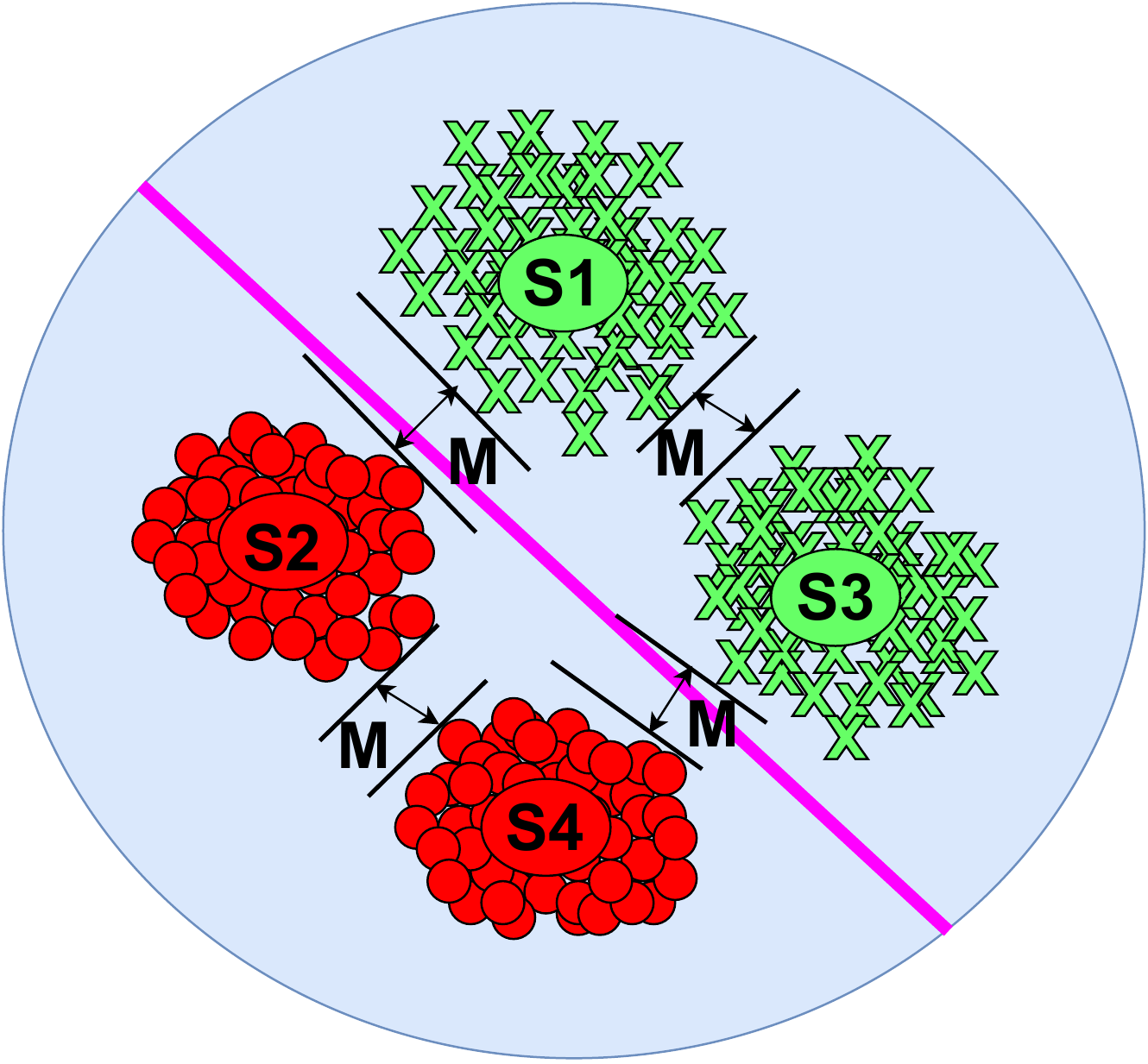}}
\subfigure[]{\label{fig:biceph_loss2}\includegraphics[width=5cm]{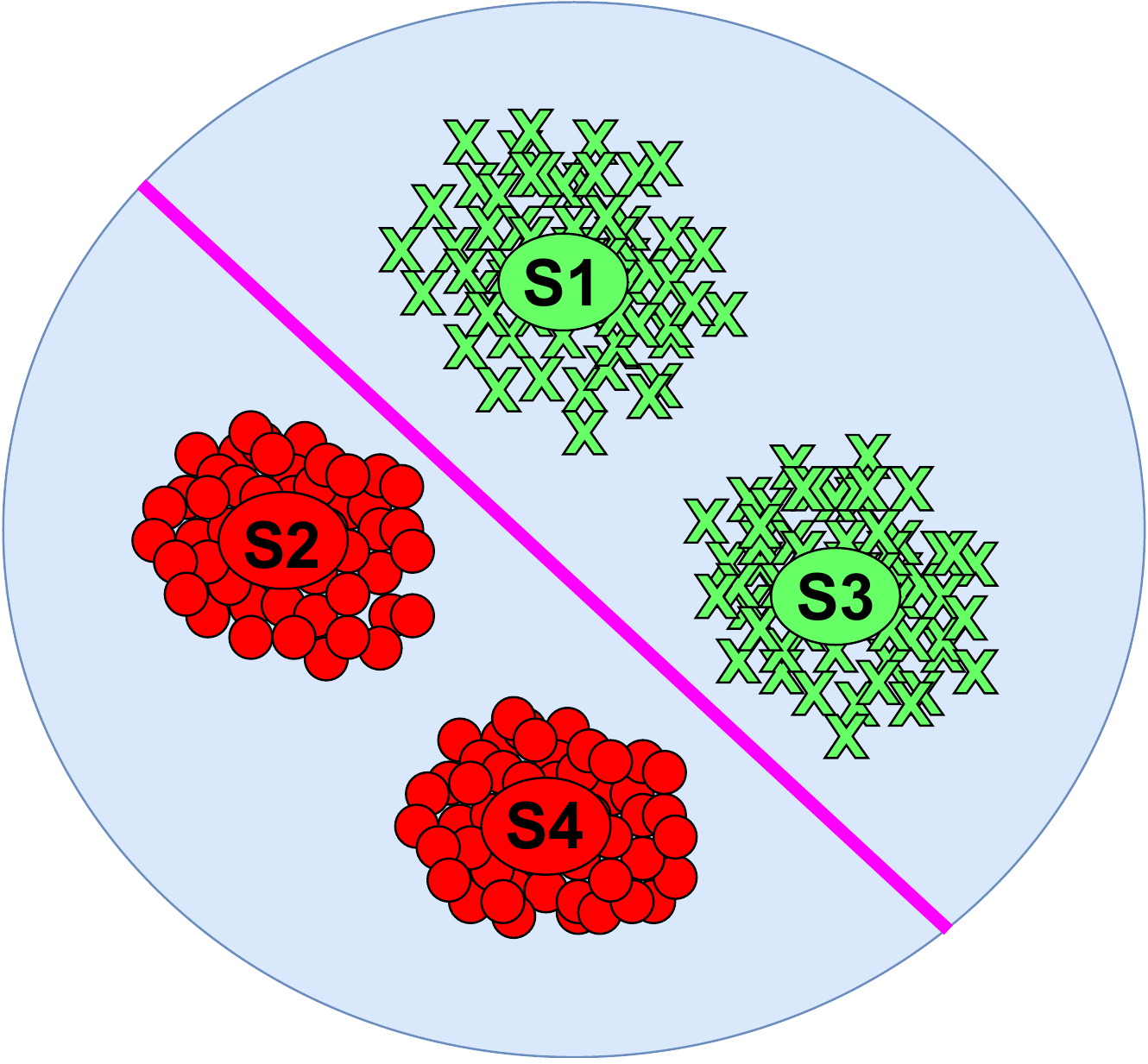}}
\caption{An intuitive representation of the embedding space after training using the Biceph-module.}
\label{spm:stats_WM3}
\end{figure*}

The CNN-RNN based methods skip this problem by also learning the inter-slice information at the cost of computational complexity. An alternative approach to counter this problem is to utilise similarity (metric) learning to keep the slices of a subject as close as possible and keep the slices of other subjects as far away as possible. Triplet loss \cite{hermans2017defense} is one of the most popular loss functions for similarity learning. Its working can be intuitively considered, as shown in \ref{fig:triplet_loss1}, as it tries to bring the slices of the same subject as close as possible while trying to keep the slices of different subjects as far away as possible. After proper training using triplet loss, the result can be visualised as shown in figure \ref{fig:triplet_loss2}. However, triplet loss does not consider the classes to which the subjects belong. As shown in figure \ref{fig:triplet_loss3}, subjects S1 and S3 belong to one class, whereas samples S2 and S4 belong to another class.

The proposed method alleviates the aforementioned issues by combining the cross-entropy loss with triplet loss. This enables the classifier to keep the slices of a subject as close as possible while keeping the slices of different subjects as far away as possible, and also create a classification boundary separating the subjects into their respective classes, as shown in figure \ref{fig:biceph_loss1}. A clearer view can also be seen in figure \ref{fig:biceph_loss2}. The major contributions of this work are as follows:
\begin{enumerate}
    \item We propose a novel framework for AD diagnosis using 2D-MRI scans and conventional 2D-CNNs.
    \item The proposed framework, used along with a VGG-16 backbone model, achieves state-of-the-art classification results on the CN vs AD, MCI vs AD and multiclass classification tasks
    \item The proposed method is computationally more efficient than the baselines while being similar in performance.
    \item The proposed framework also has an inbuilt interpretation functionality that is very helpful for understanding the decision taken by the model.
    \item The proposed framework avoids the problem of data leakage inherent in 2D-CNNs for AD diagnosis \cite{wen2020convolutional}.
\end{enumerate}

The remaining sections of this work are organised as follows:
section \ref{sec:Related Works} describes research relevant to our work, section \ref{sec:proposed} gives a detailed explanation of our proposed model, section \ref{sec:Experimentation} describes the materials and methods used in this paper and gives details of experiments performed, section \ref{sec:Results and Discussions} discusses the results obtained, and finally, section \ref{sec:Conclusion} concludes the work along with some future directions.

\vspace{4mm}
\section{Related Works}
\label{sec:Related Works}
In this section, we discuss previous research on AD classification and metric learning that are related to the proposed framework.

\vspace{3mm}
\subsection{Deep Learning Models for Alzheimer's Disease Classification}
With the growth of popularity of deep learning models, especially CNNs for image classification tasks, several models for the classification of AD were proposed, which mainly uses either 2D CNNs or 3D CNNs.

\vspace{2mm}
\subsubsection{2D Convolution Neural Networks}
 Hon et al. \cite{hon2017towards} used the method of transfer learning by fine-tuning the state-of-the-art networks, VGG16 and Inception V4, pre-trained on the ImageNet dataset for the target classification task of CN vs AD. Zhou et al. \cite{zhou2018feature} used transferred knowledge learnt from ADNI to diagnose AD samples acquired from a local hospital. Cheng et al. \cite{cheng2018robust} changed original labels to multi-bit label coding vectors and obtained features to find unrelated domains. The knowledge from unrelated domains was ignored. The usage of transferred knowledge gained from ADNI to local samples was done by Li et al. \cite{li2018detecting}. To deal with the problem of high feature dimensionality with fewer samples, Suk et al. \cite{suk2016deep} combined deep CNN with sparse regression models to diagnose AD. Sarraf et al. \cite{sarraf2017deepad} used deep CNN on functional MRI (fMRI) scans. Shi et al. \cite{shi2018multimodal} proposed stacked deep polynomial networks that leveraged complementary information from MRI and Positron Emission Tomography (PET) scans. Basaia et al. \cite{basaia2018automated} predicted MCI to AD conversion using structural MRI and deep CNN. Ieracitano et al. \cite{ieracitano2019convolutional} leveraged electroencephalography (EEG) power spectral density spectrograms and a CNN for AD diagnosis. Information from images from only one scale may not produce highly efficient results. To this end, Lu et al. \cite{lu2018multiscale} proposed a multiscale deep network that worked on Fluorodeoxyglucose Positron Emission Tomography (FDG-PET) scans.

\vspace{2mm}
\subsubsection{3D Convolution Neural Network}
To improve the classification performance of the network, some researchers used 3D-CNNs. Payan et al. \cite{payan2015predicting} found that 3D-CNN performed better in classifying CN from AD and MCI. Instead of training from scratch, which results in sub-optimal performance, Hosseini et al. \cite{hosseini2016alzheimer} used a 3D-convolutional autoencoder (CAE) to pre-train a 3D-CNN that was used for AD classification. Li et al. \cite{li2018alzheimer} also extracted information from 3D patches. A cluster of multiple dense CNNs was used on the 3D patches for AD diagnosis. Spasov et al. \cite{spasov2019parameter} used 3D separable and grouped convolutions for extracting high-level representations from Structural MRI (sMRI) for AD diagnosis. To include diversity in the model, Wang et al. \cite{wang2019ensemble} proposed an ensemble of 3D densely connected CNNs (3D-DenseNets). 

An in depth survey of many of the recent techniques proposed for AD classification can also be found in \cite{hazarika2021different, mirzaei2022machine}. 

\vspace{3mm}
\subsection{Deep Metric Learning on medical images classification}

Several works proposed the use of loss functions involving deep metric learning, which is based on the principle of similarity and dissimilarity between samples \cite{kaya2019deep}. With the advantage of data distribution, the deep metric learning loss functions can discover common patterns for each class. There are various deep metric learning techniques, such as triplet network \cite{hoffer2018deep} and Siamese network \cite{Koch2015SiameseNN}. 


Some models use multiple loss functions, both the deep metric learning loss function as well as the cross-entropy loss function for the training of their CNN model. Sun et al. \cite{Sun2016ChineseHM} used both the cross-entropy as well as the triplet loss function for the problem of recognition and retrieval of Chinese herbal medicine image, in which for the recognition task, they used the cross-entropy loss to train CNN and for the retrieval task, they fine-tuned the recognition network using triplet loss function as the images are very diverse. To overcome the problem of imbalance of images in the dataset, Lei et al. \cite{Lei2020ClassCenterIT} proposed a novel class-centre involving triplet loss for the classification of skin disease.
Huang et al. \cite{s21030764} proposed novel batch similarity-based triplet loss, which takes into account the similarity among the input images to group images of the same class and separate images of a different class, along with cross-entropy loss to improve the performance of light-weighted CNN models for the problem of medical image classification. 


\vspace{4mm}
\section{Proposed module}
\label{sec:proposed}

\begin{figure}[h]
\includegraphics[width=10cm]{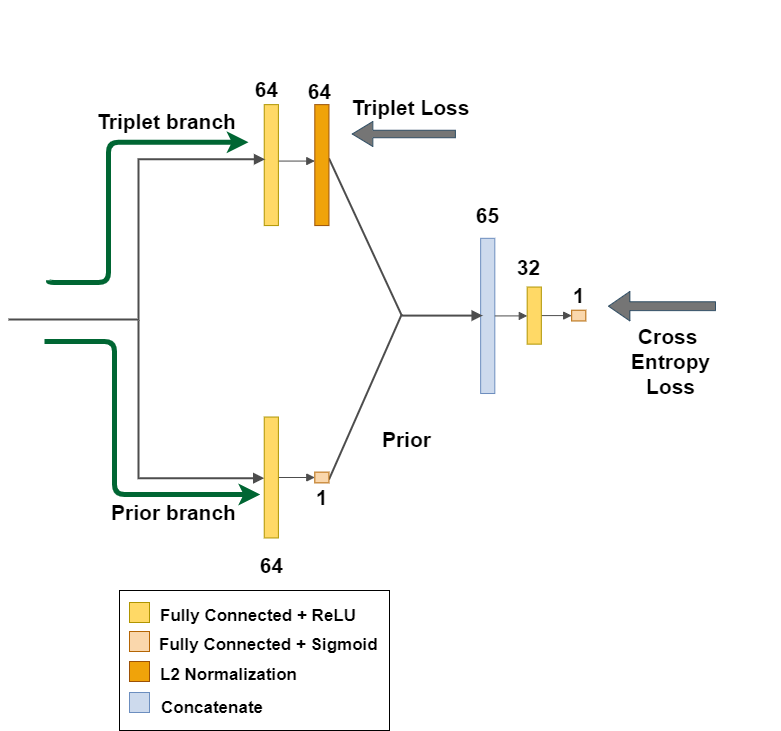}
\caption{The proposed Biceph module}
\label{biceph_module}
\end{figure}
The proposed module is shown in figure \ref{biceph_module} and is termed as the 'Biceph' module. The term 'Biceph' is inspired by the term 'Bicephalus', which means double-headed \cite{ladow2011bicephalic}. The Biceph module is placed after the flattening layer $L_{flat}$ of a 2D-CNN.The Biceph module produces a bifurcation into the triplet and the prior branches. The outputs from these two branches are appended together in the concatenate branch. The triplet branch produces an output that uses the triplet loss, whereas the concatenate branch produces an output that uses the cross-entropy loss which is responsible for the final classification of a 2D-MRI slice.

In order to train the triplet branch, we choose online semi-hard triplet mining \cite{hermans2017defense}. During the training phase, the triplet branch is trained to learn to embed 2D-MRI slices such that 2D-MRI slices of a particular subject are as close as possible, and the 2D-MRI slices of other subjects are at least separated by the mentioned margin value. The prior branch is responsible for learning a prior probability value such that the features from $L_{flat}$ belong to a specific class. The concatenate branch takes the outputs from the prior and the triplet branches to make the final classification of a slice. Thus, in the training phase, the entire Biceph module is trained to simultaneously learn:

\begin{enumerate}
    \item Embeddings in which 2D-MRI slices of a subject are placed closer to each other while being separated by a certain margin from the 2D-MRI slices of different subjects. \label{imp3}
    \item A separation boundary that separates the embeddings into different classes. \label{imp4}
\end{enumerate}

The cumulative effect of points \ref{imp3} and \ref{imp4} results in the network learning an embedding space shown in figure \ref{fig:biceph_loss2}. During the testing phase, for a 2D-MRI slice, the triplet branch gives it an embedding such that it falls into a neighbourhood that is very similar to it. This means that a 2D-MRI slice from a CN subject should be mapped to a neighbourhood with the majority of CN subjects, and a 2D-MRI slice from an AD subject should be mapped to a neighbourhood with the majority of AD subjects. 

Hence, the training process with the Biceph module can be considered such that features from layer $L_{flat}$ are given an embedding that takes them to a neighbourhood that consists of very similar nature samples. This more refined and separable embedding is then passed to a number of fully-connected layers (of concatenate branch) for final classification. Appendix materials discuss more theoretical underpinnings of the proposed Biceph module.

The value of the gradient back backpropagated through the Biceph module ($\nabla{biceph}$) can be viewed as the combination of two different terms, as shown in appendix material equation \ref{appendix_eq40}. The first term in the equation denotes the inter-slice information that the Biceph module backpropagates. On the contrary, the second term represents the intra-slice information backpropagated by the Biceph module. The inter-slice information helps perform better subject-wise classification, whereas the intra-slice information helps perform better slice-wise classification.

\begin{figure*}[h]
\centering
\includegraphics[width=1\textwidth]{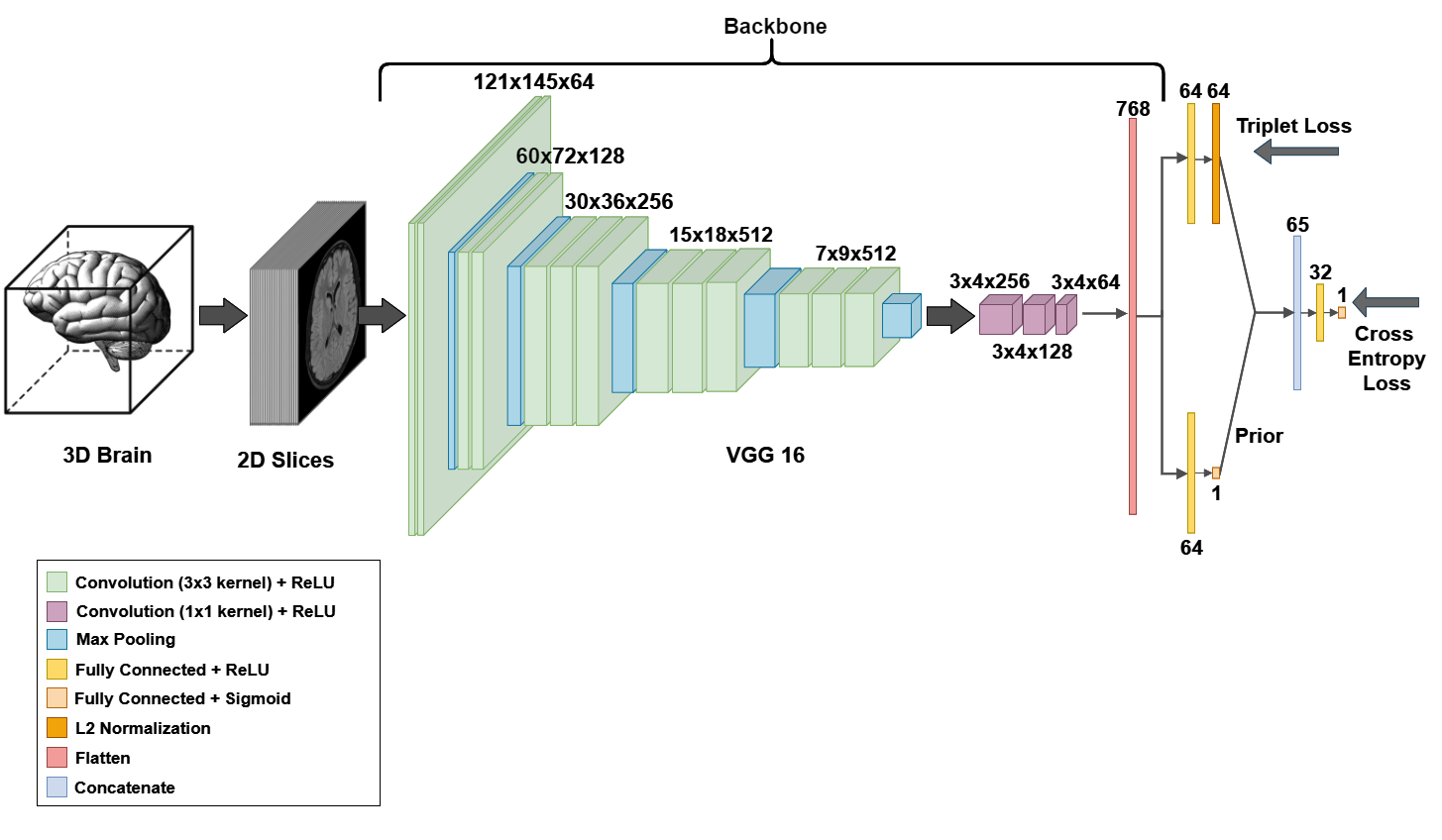}
\caption{The proposed method with backbone architecture}
\label{fig:backbone_proposed}
\end{figure*}

\vspace{4mm}
\section{Experimentation}
\label{sec:Experimentation} 
In this section, we discuss in detail about the experimental setup, dataset and the system configurations that were used in this study.

\vspace{3mm}
\subsection{System Configurations}
All the codes are implemented in the Python (version 3.7) \cite{van1995python} running on a system with Ubuntu 18.02, 256 GB RAM, Intel Xeon processor and Nvidia Titan-Xp 12 GB GPU. The deep neural network was implemented on TensorFlow backend (version 2.1) \cite{tensorflow2015-whitepaper}.
\begin{table*}
\centering
\begin{tabular}{|c|c|c|c|c|c|c|}
\hline
\multirow{2}{*}{\textbf{Models}} & \multicolumn{2}{c|}{\textbf{CN vs AD}}  & \multicolumn{2}{c|}{\textbf{MCI vs AD}} & \multicolumn{2}{c|}{\textbf{\begin{tabular}[c]{@{}c@{}}CN vs MCI \\ vs AD\end{tabular}}} \\ \cline{2-7} 
                                 & \textbf{val. acc.} & \textbf{test acc.} & \textbf{val acc.}  & \textbf{test acc.} & \textbf{val acc.}                          & \textbf{test acc.}                          \\ \hline
\multicolumn{7}{|c|}{\textbf{Axial}}     \\ \hline
VGG-Triplet \cite{hermans2017defense}             &        98.85 &                  99.63  &                \textbf{97.71}    &                  97.8  &                \textbf{98.32}    &                     97.68                                                                \\ \hline
VGG-LSTM      \cite{dua2020cnn}            &      99.77 &                 99.45   &                96.11    &                 96.52   &          96.79          &                  96.34                                                                     \\ \hline
VGG-sBiLSTM   \cite{el2020alzheimer}      &          98.17   &                  98.72  &               95.65     &                96.15    &          96.49          &        95.48                                                                               \\ \hline
Conv-LSTM     \cite{shi2015convolutional}          &     68.88    &    68.86                &             67.96       &                65.38    &           59.54         &         60.68                                                                         \\ \hline
Wang et al.   \cite{wang2018classification}          &     99.54               &     99.63               &              96.79      &        97.98            &      97.68                                      &    97.70                                         \\ \hline
\textbf{Biceph-Net}              &      \textbf{100}             &          \textbf{100}          &           96.54         &           \textbf{98.16}       &                          97.25                  &                            \textbf{97.80}                 \\ \hline
\multicolumn{7}{|c|}{\textbf{Coronal}}                                                                                                                                                                          \\ \hline
VGG-Triplet  \cite{hermans2017defense}  &              97.94      &            98.16        &            95.65        &            94.68        &              94.19                              &      96.45                                       \\ \hline
VGG-LSTM  \cite{dua2020cnn}                         &       \textbf{100}            &            \textbf{99.63}        &               97.25     &           97.62         &                             95.73               &96.34                                             \\ \hline
VGG-sBiLSTM   \cite{el2020alzheimer}                    &          \textbf{100}         &             \textbf{99.63}       &               \textbf{97.48}     &     96.70               &                     96.64                       &      95.60                                       \\ \hline
Conv-LSTM     \cite{shi2015convolutional}                   &         77.12           &           77.11         &               67.05     &       63.00             &          57.71                                  &        59.58                                     \\ \hline
Wang et al.   \cite{wang2018classification}                   &             99.54       &           \textbf{99.63}         &               97.25     &         97.25           &      \textbf{97.31}                                      &        96.94                                     \\ \hline
\textbf{Biceph-Net}              &            99.54        &         99.45           &                 96.65   &     \textbf{97.80}               &                                  96.79          &                                       \textbf{97.80}      \\ \hline
\multicolumn{7}{|c|}{\textbf{Sagittal}}                                                                                                                                                                         \\ \hline
VGG-Triplet   \cite{hermans2017defense}                   &           99.54         &        99.26            &        \textbf{97.25}            &      \textbf{98.16}              &      97.09                                      &        97.31                                     \\ \hline
VGG-LSTM    \cite{dua2020cnn}                     &        \textbf{100}            &      99.81              &      96.59              &         97.08           &         96.79                                   &                                       \textbf{97.80}      \\ \hline
VGG-sBiLSTM \cite{el2020alzheimer}                      &       \textbf{100}             &          99.63          &       96.82             &            97.63        &                                     97.25       &       96.70                                      \\ \hline
Conv-LSTM     \cite{shi2015convolutional}                     &            85.23        &      85.58              &       64.09             &          60.22          &          52.82                                  &        55.80                                     \\ \hline
Wang et al.     \cite{wang2018classification}                  &         99.54           &        99.63            &          \textbf{97.25}          &         97.98           &          \textbf{97.80}                                  &        97.40                                     \\ \hline
\textbf{Biceph-Net}              &           99.77         &     \textbf{100}               &        96.65            &       97.80             &                                 97.25           &                                     97.31        \\ \hline
\end{tabular}
\caption{Performance comparison of the baseline models and the proposed model for axial, sagittal and coronal planes.\\
{\footnotesize{
\centerline{
\textbf{Axial}: Axial plane of an MRI scan;
\textbf{Coronal}: Coronal plane of an MRI scan;
\textbf{Sagittal}: Sagittal plane of an MRI scan;} 
\centerline{
\textbf{val acc.}: Validation accuracy;
\textbf{test acc.}: Test accuracy.
}}}}

\label{biceph_results}
\end{table*}

\vspace{3mm}
\subsection{Dataset}
We acquired a total of 2500 3D-MRI scans for CN subjects, 1365 3D-MRI scans for AD subjects and 3100 3D-MRI scans for MCI subjects. The MRI scans were acquired from ADNI (1 year, 2 year and 3 year), OASIS-1, OASIS-2 and IXI databases. The entire dataset was processed using CAT-12 software. After processing raw 3D-MRI scans, we choose only Gray Matter (GM) scans. Since we focus on a balanced classification problem, we choose all 1365 GM scans from the AD class and randomly choose 1365 GM scans from MCI and CN classes. Thus, we chose 1365  3D-MRI scans from each class for the experiments performed in this paper.
\vspace{3mm}
\subsection{Experimentation Setup}
We evaluate the training, validation and testing performance of our model in the classification in three variations; CN vs AD, MCI vs AD, and CN vs MCI vs AD. We consider two types of classification: one is subject-wise classification, that is, to predict whether a particular subject has Alzheimer's or not, and the other is image-wise classification, that is, to predict whether a particular image/slice has Alzheimer's or not.

As the brain is a three-dimensional structure, so to transect the brain, it can be cut along the x, y and z planes - named the coronal plane, the axial plane and the sagittal plane as shown in the figure. The coronal plane, also known as the frontal plane or vertical plane, divides the brain into posterior and anterior portions, i.e., back and front. The axial plane, also known as the transverse plane or horizontal plane, divides the brain into the cranial and caudal portions, i.e., head and tail. The sagittal plane, also known as the lateral plane or longitudinal plane, divides the brain into the left and right portions.

For each 3D-MRI scan of a particular subject, we take 86, 112, 86 2D-slices from the middle section of the MRI scan along the Axial, Coronal and Sagittal plane, respectively. We evaluate the performance of our model in four ways:
	\begin{enumerate}
	\item  Training the model using slices along the axial plane and performing classification.
	\item  Training the model using slices along the coronal plane and performing classification.
	\item  Training the model using slices along the sagittal plane and performing classification.
	\end{enumerate}

A detailed description of the architecture and hyperparameters of the baseline models are mentioned in the appendix.

\vspace{3mm}
\subsection{Model and configuration}

 Our model uses a predefined VGG-16 network \cite{simonyan2014very} of layers with five blocks consisting of several 2D convolution layers with a kernel of size 3x3 and a max-pooling layer. These layers are followed by three sequential layers of 2D convolution layer, each with a kernel of size 1x1, which helps in the reduction of dimensions. All the convolution layers also include the use of rectified linear activation function. The results from these convolution layers are flattened into one-dimension from the three-dimension input using a Flatten layer.

The flattened features are then passed onto two different branches: (a) triplet branch and (b) prior branch. In (a), the flattened features are passed onto a $64\, \times \, 1$-D dense layer, which is passed to the triplet loss layer after doing an L2-normalization. The prior branch (b) is responsible for providing prior knowledge regarding the class of the flattened features. In (b), the flattened features are passed onto a dense layer with one output that consists of a sigmoid activation function for binary classification tasks. For multiclass classification tasks, (b) consists of a $3\,\times\,1$-D dense layer with the softmax activation function. The branches (a) and (b) are then combined and passed onto three more dense layers. The final output from these layers is sent to the cross-entropy loss layer. The algorithm for online triplet mining and loss calculation for Biceph-Net is given in appendix material algorithm 1.


The batch size is taken as 80, with each batch having a uniform distribution of samples/images. The learning rate is initialised with 0.001 and reduced using ReduceLROnPlateau, i.e., the learning rate is reduced by the factor of 0.1 when the metric, here validation loss, has stopped improving. The number of epochs is 100. We split the dataset into 80\% training and 20\% testing set. We further split the training dataset to 80\% training and 20\% validation.

\vspace{4mm}
\section{Results and Discussions}
\label{sec:Results and Discussions}
In this section, we present and discuss the the classification results achieved by each model covered in this study. We also delve into discussions on the computational complexity of the models, feature embeddings and the interpretation of decision taken by the proposed model.



\vspace{3mm}
\subsection{Comparison of classification performance}
{\textbf{CN vs AD: }}We begin with the simplest of classification tasks in AD diagnosis, CN vs AD. A comparative analysis of the performances for CN vs AD classification task on axial, coronal and sagittal planes are given in Table \ref{biceph_results}. We can observe that the proposed Biceph-Net performs better than CNN-RNN models for axial and sagittal planes. In the coronal plane, the performance of Biceph-Net is almost similar to that of VGG-LSTM and VGG-sBiLSTM. We can also observe that the straightforward eight-layer model proposed by Wang et al. \cite{wang2018classification} also performs equally to Biceph-Net in all these cases. It is also important to note that Biceph-Net and eight-layer CNN consists of entirely different structures, and such a straightforward comparison might not be fair. In order to address this, we also implemented the Biceph version of eight-layer CNN, in which we introduced our Biceph module after the flattening layer. This Biceph-eight-layer model performed better than vanilla eight-layer CNN and was more computationally efficient. We can also observe that Biceph-Net outperforms the VGG-Triplet network in all three scenarios.

The proposed network easily outperforms other baseline methods on the Axial and Sagittal planes. However, in the coronal plane, it misses out by inducing some misclassifications. The CN vs AD classification task is considered the simplest of tasks for AD diagnosis because the MRI scans of AD subjects show highly visible neurodegeneration compared to that of the CN subjects. This makes the networks learn the differences between images of the two classes with much ease. The proposed method can be used with axial and sagittal slices to provide a robust classification for CN vs AD in the real world scenario if someone decides to use it. The CN vs AD classification is found to produce very high classification accuracies \cite{tanveer2020machine}. We can also observe that the proposed network from Table \ref{biceph_results} performed CN vs AD classification with 100\% accuracy in the Axial and Sagittal planes.

{\textbf{MCI vs AD: }} We also perform a comparative analysis of all the models on the more complex task of differentiating between MCI and AD subjects. The results for axial, coronal and sagittal planes are given in Table \ref{biceph_results}. Similar to the CN vs AD classification scenario, the Biceph-Net performs equally or better than more complex CNN-RNN models on all three planes. For axial and sagittal planes, Biceph-Net outperforms CNN-RNN models. For the coronal plane, its performance is very similar to that of VGG-LSTM. Biceph-Net outperforms eight-layer CNN for axial and coronal planes.

MCI vs AD is considered a much more challenging classification task than the CN vs AD task as the MRI scans of MCI and AD subjects can visually be very similar \cite{basaia2018automated}. This is due to the fact that the MCI subjects in later stages can be very similar to the early stage AD subjects. Therefore, this poses a much more challenging task for vanilla CNNs as they might classify most of the slices of a subject to a different class, causing the subject's misclassification. However, Biceph-Net can also deal with this scenario and outperforms all the baseline methods for the Axial and Coronal planes. It also performs almost as good as much more computationally intensive baseline methods as it makes one extra misclassification than the baseline methods.   Thus, if it is being deployed in the real world, then Biceph-Net can be used with the Axial and coronal planes for MCI vs AD classification.

{\textbf{CN vs MCI vs AD: }} We also perform a comparative analysis on the challenging problem of multiclass classification. The problem becomes challenging because the non-converter MCI subjects are more similar to CN subjects, whereas the converter-MCI subjects are more similar to AD subjects \cite{zhang2011multimodal}. The results for axial, coronal and sagittal planes are given in Table \ref{biceph_results}. We can observe that Biceph-Net also performs equally well in multiclass classification. Its performance is similar or better than the CNN-RNN models. In this case, Biceph-Net implemented upon the axial and coronal slices outperform the baseline methods. Thus, even for much more complicated tasks like multiclass classification (CN vs MCI vs AD) and MCI vs AD binary classification tasks, the Biceph-Net produces state of the art classification results.

\vspace{3mm}
\subsection{Comparison of computational complexity}
In this section, we compare the computational complexity of the models given in Table \ref{biceph_results}. In order to compare the computational complexity, we choose three modes of comparison: (a) Total number of parameters in millions (MParams), (b) Size of the model on disk in megabytes (MB), (c) The total number of Floating Point Operations Per Second (FLOPS) of each model in millions (MFlops). In order to perform this comparison, we choose all the models trained on the axial slices of the MCI vs AD classification task.

The total number of trainable parameters in ConvLSTM is 15.402 MParams, VGG-LSTM is 15.093 MParams, VGG-sBiLSTM is 15.361 MParams, Triplet network is 14.936 MParams, and Biceph-Net is 14.987 MParams, respectively. The disk space occupied by each model is as follows:  ConvLSTM occupies 61 MB, VGG-LSTM occupies 60 MB, VGG-sBiLSTM occupies 65 MB, whereas both Triplet network and Biceph-Net occupy 58 MB each. We can observe that the Biceph-Net has a lesser number of trainable parameters than ConvLSTM, VGG-LSTM, and VGG-sBiLSTM. As a result, it also occupies less space on the disk. Thus, Biceph-net also has a computational advantage over these more computationally intensive baseline models. Since it also occupies less space on disk, it is also more efficient than the CNN-RNN baseline models in being deployed on lightweight edge devices.

The total flops consumed by each model is as follows: (a) ConvLSTM – 30.941 MFlops, VGG-LSTM – 30.198 MFlops, VGG-sBiLSTM - 30.791 MFlops, Triplet - 29.863 MFlops, Biceph-Net - 29.965 MFlops. The Biceph-Net has an almost marginally higher FLOPS count with the Triplet-network while being much superior in performance. However, it saves almost 1 MFlops compared to the baseline CNN-RNN architectures. Another thing to observe is that the Biceph-module can be used with any other backbone network. Hence, it is possible that a much lighter backbone along with the Biceph-module produces a similar performance as to that of the models in Table \ref{biceph_results}. We keep further exploration on this as future work.

We further perform a paired t-test on the classification results obtained in Table \ref{biceph_results} with a significance value of 0.05. We get a p-value of 0.073 for CN vs AD, 0.056 for MCI vs AD and 0.057 for CN vs MCI vs AD. Thus, we cannot reject the null hypothesis that the classification results of the baseline and proposed methods are similar. This also emphasises the fact that the performance of Biceph-Net is similar to the baselines while being more computationally efficient.

\vspace{3mm}
\subsection{Visualisation of feature embeddings}
We also visualise the feature embeddings learnt by VGG-Triplet and Biceph networks, respectively. The visualisation is done using PCA, and T-sne \cite{van2008visualizing} plots. For visualising the feature embedding learnt by Biceph-Net, we visualise the 64-dimensional feature vector produced by the triplet branch.
We choose 10,000 randomly chosen feature vectors from the test set of axial slices for plotting. For the VGG-Triplet network, the PCA plots for CN vs AD and MCI vs AD are given in appendix figures \ref{fig:pca}(a)  and \ref{fig:pca}(b). We can observe that samples from two different classes are in different clusters, and there is a high entanglement among the clusters. On the other hand, for the Biceph-Net, the embeddings are shown in appendix figures \ref{fig:pca}(d) and \ref{fig:pca}(e). We observe a better separation of the samples of two classes for Biceph-Net. This also helps in visually verifying that the embeddings learnt by Biceph-Net are roughly similar to that shown in figure \ref{fig:biceph_loss2}. 

A more fine-grained non-linear relationship can be visually verified by appendix figure \ref{tsne_biceph}. From appendix figure \ref{tsne_biceph}(a) and \ref{tsne_biceph}(b), we can observe that for the VGG-Triplet network, the samples are grouped in different clusters so that a cluster of one class can have its neighbourhood occupied by a cluster of the other class. Using a $K$-NN classifier on such an embedding will result in more misclassifications as the majority of the neighbours can belong to a different class. However, in appendix figures \ref{tsne_biceph}(d) and \ref{tsne_biceph}(e), the Biceph-Net produces more distinct neighbourhoods for the clusters of samples. We can observe that the clusters of samples of the same class are much closer and clusters of samples of different classes have a larger degree of separability. This nature of embedding helps in more accurate classification, even for a challenging task like MCI vs AD.

For multiclass classification, the PCA plots can be seen in appendix figures \ref{fig:pca}(c) and \ref{fig:pca}(f) for VGG-Triplet and Biceph-Net, respectively. The three classes are represented in blue, red and pink colours. We can observe that the neighbourhoods for the three classes are more clearly visible in Biceph-Net as compared to its counterpart. A similar trend can also be found in the appendix figures \ref{tsne_biceph}(c) and \ref{tsne_biceph}(f), wherein the embedding of Biceph-Net shows more distinct neighbourhoods of samples of three different classes. It also shows that the samples of different classes have lesser intermixing between them, thus, leading to a better classification result than VGG-Triplet with $K$-NN classifier.

\vspace{3mm}
\subsection{Interpreting model decision}
The rich neighbourhood structure learnt by the embedding in the triplet branch can be utilised to interpret the classification decision taken by the network. We observed that whenever a 2D slice is correctly classified, the neighbourhood of its embedding is marked by a majority of samples belonging to the same class. Whenever a 2D slice is misclassified, then the neighbourhood of its embedding contains a majority of samples of another class. Owing to this observation, we performed an analysis of classifications of six randomly chosen test subjects from CN vs AD coronal experiments. Out of the six subjects, three were correctly classified, and three were misclassified.

For the misclassified subjects, the number of slices and the class of their corresponding majority $K$ nearest neighbours are given in Table \ref{biceph_embbedding_knn}. We can observe that for all these subjects, the majority of the $K$ nearest neighbours belong to the opposite class. A similar analysis was performed for the three correctly classified samples, and we found that for all slices, their respective nearest neighbours belonged to the same class. 

Hence, investigating the neighbourhood could provide an additional nearest neighbour based interpretation of the classification decision produced by Biceph-Net.

\begin{table}[h]
\tiny
\begin{tabular}{|c|c|c|c|c|c|c|c|c|c|c|c|}
\hline
\multirow{3}{*}{\textbf{Subject ID}} & \multirow{3}{*}{\textbf{\begin{tabular}[c]{@{}c@{}}True\\ label\end{tabular}}} & \multicolumn{10}{c|}{\textbf{Values of $K$}}                                                                                                                                             \\ \cline{3-12} 
                                     &                                                                                & \multicolumn{2}{c|}{\textbf{$K=10$}} & \multicolumn{2}{c|}{\textbf{$K=20$}} & \multicolumn{2}{c|}{\textbf{$K=30$}} & \multicolumn{2}{c|}{\textbf{$K=40$}} & \multicolumn{2}{c|}{\textbf{$K=50$}} \\ \cline{3-12} 
                                     &                                                                                & \textbf{CN}      & \textbf{AD}     & \textbf{CN}      & \textbf{AD}     & \textbf{CN}      & \textbf{AD}     & \textbf{CN}      & \textbf{AD}     & \textbf{CN}      & \textbf{AD}     \\ \hline
372                                  & AD                                                                             & 72               & 40              & 71               & 41              & 70               & 42              & 69               & 43              & 69               & 43              \\ \hline
528                                  & AD                                                                             & 58               & 54              & 56               & 56              & 54               & 58              & 54               & 58              & 54               & 58              \\ \hline
543                                  & AD                                                                             & 62               & 50              & 63               & 49              & 63               & 49              & 62               & 50              & 62               & 50              \\ \hline
\end{tabular}
\caption{Analysis of misclassified subjects for CN vs AD coronal slices. $K$ represents the number of neighbours taken into consideration.}
\label{biceph_embbedding_knn}
\end{table}

\vspace{4mm}
\section{Conclusions and future works}
\label{sec:Conclusion}
In this work, we presented a novel `Biceph-module' that is robust and computationally efficient for diagnosis of Alzheimer's disease using 2D-MRI scans. The module can be attached to any 2D-CNN by placing it after the flattening layer. We presented such an architecture with the VGG-16 backbone network termed as `Biceph-Net'. Biceph-Net outperformed baseline models in most of the experiments. Biceph-Net was also able to achieve equivalent performances with many complicated CNN-RNN models. We also visualise the feature embeddings of Biceph-Net with that of triplet network and gain an intuitive understanding as to why Biceph-Net is better as compared to the baselines. From this work, we can conclude that Biceph-Net is computationally more efficient while being similar in performance to the baseline models. Thus, it can be a better alternative for AD diagnosis as compared to the baseline models.

The `Biceph-module' can also be used with other 2D-CNN backbones for disease diagnosis using 2D-MRI scans. It can also be used in scenarios where the data is rendered in a hierarchical format as explained in section \ref{introduction}. A limitation of the proposed framework is that it works on only a single feature view and does not consider multiple feature views. Another limitation of the proposed framework is that it doesn't have any inherent mechanism to ignore outlier embeddings which may adversely affect the final classification.

In the future, we plan to explore the rich neighbourhood structure that the Biceph-module provides for further interpreting the classifications of the network. We also plan to utilise the low dimension feature embeddings for amalgamating the different feature views obtained from axial, coronal and sagittal planes. We also plan to work on reducing the effect of outlier embeddings in order to produce more robust classifications.

\vspace{4mm}
\section{Data Availability Statement}
All the codes and data can be utilized in this paper can be found in the following Github repository: \url{https://github.com/mtanveer1}.


%



\vspace{4mm}
\section*{Acknowledgment}
This work is supported by National Supercomputing Mission under DST and Miety, Govt. of India under Grant No.
DST/NSM/R\&D HPC Appl/2021/03.29. This work is also supported by Department of Science and Technology, INDIA under Ramanujan fellowship scheme grant no. SB/S2/RJN-001/2016, Council of Scientific \& Industrial Research (CSIR), New Delhi, INDIA under Extra Mural Research (EMR) scheme grant no. 22(0751)/17/EMR-II, and Science and Engineering Research Board (SERB), INDIA under Early Career Research Award scheme grant no. ECR/2017/000053. We also thank
the Indian Institute of Technology Indore for the support and
facility offered for the research.

\bibliographystyle{IEEEtran}
\bibliography{bibliography}




%

\begin{appendices}

\setcounter{table}{0}
\renewcommand{\thetable}{A\arabic{table}}
\setcounter{figure}{0} 
\renewcommand\thefigure{A\arabic{figure}}    
\section{ Description of the model }
In this section, we give a brief description about the theoretical underpinnings of the proposed model. Let $T_r$, $P$ and $C$ denote the triplet branch, prior branch and concatenate branch respectively. Let $n_{T}$, $n_P$, and $n_C$ be the total number of fully connected (fc) layers in $T_r$, $P$ and $C$ respectively. For a fully connected layer $l$, the forward pass equation can be written in the following steps:
\begin{align}
a &= W.X,  \label{eq1}\\
o &= f(a), \label{eq2}
\end{align}
where, $X$ represents the input  feature vector (including the numeral 1), $W$ represents the parameter matrix including the bias term, $f(.)$ represents the activation function and $o$ denotes the output from $l$ during the forward pass.

Let $l_{0}^{T_r},l_{1}^{T_r},...,l_{n_{T}}^{T_r}$ denote the fc layers of $T_r$. Similarily, $l_{0}^{P},l_{1}^{P},...,l_{n_{P}}^{P}$ denote fc layers of $P$ and $l_{0}^{C},l_{1}^{C},...,l_{n_{C}}^{C}$ denote fc layers of $C$.

\begin{table*}
\centering
\begin{tabular}{|c|c|c|c|c|c|c|c|c|}
\hline
\textbf{Task}                    & \textbf{Model} & \textbf{Plane}            & \textbf{\begin{tabular}[c]{@{}c@{}}Sequence \\ size\end{tabular}} & \textbf{Learning Rate (LR)}                & \textbf{\begin{tabular}[c]{@{}c@{}}Batch\\ size\end{tabular}} & \textbf{\begin{tabular}[c]{@{}c@{}}LR \\ schedule\end{tabular}}                              & \textbf{Patience}   & \multicolumn{1}{l|}{\textbf{Epochs}} \\ \hline
\multirow{9}{*}{CN vs AD}        & VGG-LSTM       & \multirow{3}{*}{axial}    & \multirow{3}{*}{86}                                               & \multirow{3}{*}{$10^{-6}$}  & \multirow{3}{*}{4}                                            & \multirow{27}{*}{\begin{tabular}[c]{@{}c@{}}Reduce\\  LR on \\ plateau\\ (0.1)\end{tabular}} & \multirow{3}{*}{5}  & \multirow{3}{*}{40}                  \\ \cline{2-2}
                                 & VGG-sBiLSTM    &                           &                                                                   &                            &                                                               &                                                                                              &                     &                                      \\ \cline{2-2}
                                 & Conv-LSTM      &                           &                                                                   &                            &                                                               &                                                                                              &                     &                                      \\ \cline{2-6} \cline{8-9} 
                                 & VGG-LSTM       & \multirow{3}{*}{coronal}  & \multirow{3}{*}{112}                                              & \multirow{3}{*}{$10^{-5}$}   & \multirow{3}{*}{1}                                            &                                                                                              & \multirow{3}{*}{5}  & \multirow{3}{*}{35}                  \\ \cline{2-2}
                                 & VGG-sBiLSTM    &                           &                                                                   &                            &                                                               &                                                                                              &                     &                                      \\ \cline{2-2}
                                 & Conv-LSTM      &                           &                                                                   &                            &                                                               &                                                                                              &                     &                                      \\ \cline{2-6} \cline{8-9} 
                                 & VGG-LSTM       & \multirow{3}{*}{sagittal} & \multirow{3}{*}{86}                                               & \multirow{2}{*}{$10^{-4}$}    & \multirow{2}{*}{4}                                            &                                                                                              & \multirow{2}{*}{2}  & \multirow{2}{*}{35}                  \\ \cline{2-2}
                                 & VGG-sBiLSTM    &                           &                                                                   &                            &                                                               &                                                                                              &                     &                                      \\ \cline{2-2} \cline{5-6} \cline{8-9} 
                                 & Conv-LSTM      &                           &                                                                   & $10^{-5}$                    & 1                                                             &                                                                                              & 5                   & 35                                   \\ \cline{1-6} \cline{8-9} 
\multirow{9}{*}{MCI vs AD}       & VGG-LSTM       & \multirow{3}{*}{axial}    & \multirow{3}{*}{86}                                               & $10^{-5}$                   & \multirow{3}{*}{4}                                            &                                                                                              & \multirow{3}{*}{5}  & 30                                   \\ \cline{2-2} \cline{5-5} \cline{9-9} 
                                 & VGG-sBiLSTM    &                           &                                                                   & \multirow{2}{*}{$10^{-6}$}  &                                                               &                                                                                              &                     & 40                                   \\ \cline{2-2} \cline{9-9} 
                                 & Conv-LSTM      &                           &                                                                   &                            &                                                               &                                                                                              &                     & 50                                   \\ \cline{2-6} \cline{8-9} 
                                 & VGG-LSTM       & \multirow{3}{*}{coronal}  & \multirow{3}{*}{112}                                              & \multirow{15}{*}{$10^{-6}$} & 1                                                             &                                                                                              & \multirow{15}{*}{5} & 30                                   \\ \cline{2-2} \cline{6-6} \cline{9-9} 
                                 & VGG-sBiLSTM    &                           &                                                                   &                            & \multirow{8}{*}{4}                                            &                                                                                              &                     & 40                                   \\ \cline{2-2} \cline{9-9} 
                                 & Conv-LSTM      &                           &                                                                   &                            &                                                               &                                                                                              &                     & 50                                   \\ \cline{2-4} \cline{9-9} 
                                 & VGG-LSTM       & \multirow{3}{*}{sagittal} & \multirow{3}{*}{86}                                               &                            &                                                               &                                                                                              &                     & 30                                   \\ \cline{2-2} \cline{9-9} 
                                 & VGG-sBiLSTM    &                           &                                                                   &                            &                                                               &                                                                                              &                     & \multirow{2}{*}{50}                  \\ \cline{2-2}
                                 & Conv-LSTM      &                           &                                                                   &                            &                                                               &                                                                                              &                     &                                      \\ \cline{1-4} \cline{9-9} 
\multirow{9}{*}{CN vs MCI vs AD} & VGG-LSTM       & \multirow{3}{*}{axial}    & \multirow{3}{*}{86}                                               &                            &                                                               &                                                                                              &                     & 30                                   \\ \cline{2-2} \cline{9-9} 
                                 & VGG-sBiLSTM    &                           &                                                                   &                            &                                                               &                                                                                              &                     & \multirow{2}{*}{50}                  \\ \cline{2-2}
                                 & Conv-LSTM      &                           &                                                                   &                            &                                                               &                                                                                              &                     &                                      \\ \cline{2-4} \cline{6-6} \cline{9-9} 
                                 & VGG-LSTM       & \multirow{3}{*}{coronal}  & \multirow{3}{*}{112}                                              &                            & 1                                                             &                                                                                              &                     & 30                                   \\ \cline{2-2} \cline{6-6} \cline{9-9} 
                                 & VGG-sBiLSTM    &                           &                                                                   &                            & \multirow{5}{*}{4}                                            &                                                                                              &                     & \multirow{2}{*}{50}                  \\ \cline{2-2}
                                 & Conv-LSTM      &                           &                                                                   &                            &                                                               &                                                                                              &                     &                                      \\ \cline{2-4} \cline{9-9} 
                                 & VGG-LSTM       & \multirow{3}{*}{sagittal} & \multirow{3}{*}{86}                                               &                            &                                                               &                                                                                              &                     & 30                                   \\ \cline{2-2} \cline{9-9} 
                                 & VGG-sBiLSTM    &                           &                                                                   &                            &                                                               &                                                                                              &                     & \multirow{2}{*}{50}                  \\ \cline{2-2}
                                 & Conv-LSTM      &                           &                                                                   &                            &                                                               &                                                                                              &                     &                                      \\ \hline
\end{tabular}

\caption{Hyperparameter configurations for baseline models used in this paper}
\label{model_config}
\end{table*}

The forward pass equations for the Biceph module are as follows:

\begin{enumerate}

    \item \textbf{Concatenate branch:}

 \begin{align}
    \underline{\text{for }l_{0}^{C}}\text{: } & \notag \\
        a_0^{C} &= W_0^{C}X^{C} \,,\\
        o_0^{C} &= f_0^{C}(a_0^{C}) \,, \\
        \text{similarly, }\underline{\text{for }l_{1}^{C}}\text{: } & \notag \\
        a_1^{C} &= W_1^{C}o_0^{C}, \\
        o_1^{C} &= f_1^{C}(a_1^{C}), \\
    &\vdots\\
    \underline{\text{for }l_{n_C}^{C}}\text{: } & \notag \\
        a_{n_C}^{C} &= W_{n_C}^{C}o_{n_C-1}^{C},  \\
        o_{n_C}^{C} &= f_{n_C}^{C}(a_{n_C}^{C}),
  \end{align}
    
    \item \textbf{Triplet branch:}
    \begin{align}
    \underline{\text{for }l_{0}^{T_r}}\text{: } & \notag \\
        a_0^{T_r} &= W_0^{T_r}X^{T_r} \,,\\
        o_0^{T_r} &= f_0^{T_r}(a_0^{T_r}) \,, \\
        \text{similarly, }\underline{\text{for }l_{1}^{T_r}}\text{: } & \notag \\
        a_1^{T_r} &= W_1^{T_r}o_0^{T_r}, \\
        o_1^{T_r} &= f_1^{T_r}(a_1^{T_r}), \\
    &\vdots\\
    \underline{\text{for }l_{n_T}^{T_r}}\text{: } & \notag \\
        a_{n_T}^{T_r} &= W_{n_T}^{T_r}o_{n_T-1}^{T_r},  \\
        o_{n_T}^{T_r} &= f_{n_T}^{T_r}(a_{n_T}^{T_r}),
  \end{align}
    \item \textbf{Prior branch:}
    \begin{align}
    \underline{\text{for }l_{0}^P}\text{: } & \notag \\
        a_0^P &= W_0^PX^P \,,\\
        o_0^{P} &= f_0^{P}(a_0^{P}) \,, \\
        \text{similarly, }\underline{\text{for }l_{1}^{P}}\text{: } & \notag \\
        a_1^{P} &= W_1^{P}o_0^{P}, \\
        o_1^{P} &= f_1^{P}(a_1^{P}), \\
    &\vdots\\
    \underline{\text{for }l_{n_P}^{P}}\text{: } & \notag \\
        a_{n_P}^{P} &= W_{n_P}^{P}o_{n_P-1}^{P},  \\
        o_{n_P}^{P} &= f_{n_P}^{P}(a_{n_T}^{P}),
  \end{align}
\end{enumerate}

The backward pass equations for the biceph module are as follows:
\begin{enumerate}
    \item \textbf{Concatenate branch:}
      \begin{align}
    \underline{\text{for }l_{n_C}^{C}}\text{: } & \notag \\
    \frac{\partial C_E}{\partial l_{n_C}^{C}} &= \begin{bmatrix} \frac{\partial C_E}{\partial o_{n_C}^C} \frac{\partial o_{n_C}^C}{\partial f_{n_C}^C}\frac{\partial f_{n_C}^C}{\partial a_{n_C}^C}\frac{\partial a_{n_C}^C}{\partial W_{n_C}^C}\end{bmatrix}, 
    \end{align}
   
      \begin{multline*}
    \underline{\text{for }l_{n_C-1}^{C}}\text{: }  \notag \\
        \frac{\partial C_E}{\partial l_{n_C-1}^{C}} = \biggl[\frac{\partial C_E}{\partial o_{n_C}^C} \frac{\partial o_{n_C}^C}{\partial f_{n_C}^C}\frac{\partial f_{n_C}^C}{\partial a_{n_C}^C}\frac{\partial a_{n_C}^C}{\partial o_{n_C-1}^C}\\
    \frac{\partial o_{n_C-1}^C}{\partial f_{n_C-1}^C}\frac{\partial f_{n_C-1}^C}{\partial a_{n_C-1}^C}\frac{\partial a_{n_C-1}^C}{\partial W_{n_C-1}^C}\biggr] \tag{25},\\
        \end{multline*} 
   \vspace{-10mm} 
    \begin{align}
        \vdots \notag
    \end{align}
    \vspace{-8mm} 
     \begin{multline*}
    \underline{\text{for }l_{0}^{C}}\text{: }  \notag \\
        \frac{\partial C_E}{\partial l_{0}^{C}} = \biggl[\frac{\partial C_E}{\partial o_{n_C}^C} \frac{\partial o_{n_C}^C}{\partial f_{n_C}^C}\frac{\partial f_{n_C}^C}{\partial a_{n_C}^C}\frac{\partial a_{n_C}^C}{\partial o_{n_C-1}^C}\\
    \frac{\partial o_{n_C-1}^C}{\partial f_{n_C-1}^C}\frac{\partial f_{n_C-1}^C}{\partial a_{n_C-1}^C}\frac{\partial a_{n_C-1}^C}{\partial o_{n_C-2}^C}\hdots \\
    \frac{\partial o_{n1}^C}{\partial   f_{n1}^C}\frac{\partial f_{n1}^C}{\partial a_{n1}^C}\frac{\partial a_{n1}^C}{\partial o_{n_0}^C}\frac{\partial o_{n_0}^C}{\partial f_{n_0}^C}\frac{\partial f_{n_0}^C}{\partial a_{n_0}^C}\frac{\partial a_{n_0}^C}{\partial W_{n_0}^C}
    \biggr] \tag{26},
    \end{multline*} 
 Let $Q$ = 
    \begin{multline*}
    \biggl[\frac{\partial C_E}{\partial o_{n_C}^C} \frac{\partial o_{n_C}^C}{\partial f_{n_C}^C}\frac{\partial f_{n_C}^C}{\partial a_{n_C}^C}\frac{\partial a_{n_C}^C}{\partial o_{n_C-1}^C}\\
    \frac{\partial o_{n_C-1}^C}{\partial f_{n_C-1}^C}\frac{\partial f_{n_C-1}^C}{\partial a_{n_C-1}^C}\frac{\partial a_{n_C-1}^C}{\partial o_{n_C-2}^C}\hdots \\
    \frac{\partial o_{n1}^C}{\partial   f_{n1}^C}\frac{\partial f_{n1}^C}{\partial a_{n1}^C}\frac{\partial a_{n1}^C}{\partial o_{n_0}^C}\frac{\partial o_{n_0}^C}{\partial f_{n_0}^C}\frac{\partial f_{n_0}^C}{\partial a_{n_0}^C}\biggr] \tag{27},
    \end{multline*} 
    
    \item \textbf{Triplet branch:}
    The triplet branch receives one gradient from the triplet loss and another gradient from the concatenate branch.

    \begin{align}
    \underline{\text{for }l_{n_T}^{T_r}}\text{: } &  \notag\\
    \nabla l_{n_T}^{T_r} &= \frac{\partial T_{rp}}{\partial l_{n_T}^{T_r}} + \frac{\partial C_E}{\partial l_{n_T}^{T_r}},\tag{28} \\
    &= \biggl[\frac{\partial T_{rp}}{\partial o_{n_T}^{T_r}}\frac{\partial o_{n_T}^{T_r}}{\partial f_{n_T}^{T_r}}\frac{\partial f_{n_T}^{T_r}}{\partial a_{n_T}^{T_r}}\frac{\partial a_{n_T}^{T_r}}{\partial W_{n_T}^{T_r}}\biggr] \notag\\
    & + \biggl[Q \frac{\partial f_{n_0}^{C}}{\partial o_{n_T}^{T_r}} \frac{\partial o_{n_T}^{T_r}}{\partial f_{n_T}^{T_r}}\frac{\partial f_{n_T}^{T_r}}{\partial a_{n_T}^{T_r}}\frac{\partial a_{n_T}^{T_r}}{\partial W_{n_T}^{T_r}} \biggr], \tag{29}
    \end{align}
    
    \begin{align}
      \underline{\text{for }l_{n_T-1}^{T_r}}\text{: } & \notag \\
        \nabla l_{n_T-1}^{T_r} &= \frac{\partial T_{rp}}{\partial l_{n_T-1}^{T_r}} + \frac{\partial C_E}{\partial l_{n_T-1}^{T_r}}, \tag{30} 
    \end{align}
  
    \begin{multline*}
 \nabla l_{n_T-1}^{T_r} =  \biggl[\frac{\partial T_{rp}}{\partial o_{n_T}^{T_r}}\frac{\partial o_{n_T}^{T_r}}{\partial f_{n_T}^{T_r}}\frac{\partial f_{n_T}^{T_r}}{\partial a_{n_T}^{T_r}}\frac{\partial a_{n_T}^{T_r}}{\partial o_{n_T-1}^{T_r}} \notag\\
  \frac{\partial o_{n_T-1}^{T_r}}{\partial f_{n_T-1}^{T_r}}\frac{\partial f_{n_T-1}^{T_r}}{\partial a_{n_T-1}^{T_r}}\frac{\partial a_{n_T-1}^{T_r}}{\partial W_{n_T-1}^{T_r}}\biggr] \notag\\
  + \biggl[Q \frac{\partial f_{n_0}^{C}}{\partial o_{n_T}^{T_r}} \frac{\partial o_{n_T}^{T_r}}{\partial f_{n_T}^{T_r}}\frac{\partial f_{n_T}^{T_r}}{\partial a_{n_T}^{T_r}}\frac{\partial a_{n_T}^{T_r}}{\partial o_{n_T-1}^{T_r}} \notag\\
  \frac{\partial o_{n_T-1}^{T_r}}{\partial f_{n_T-1}^{T_r}}\frac{\partial f_{n_T-1}^{T_r}}{\partial a_{n_T-1}^{T_r}}\frac{\partial a_{n_T-1}^{T_r}}{\partial W_{n_T-1}^{T_r}}\biggr], \tag{31}
    \end{multline*} 
    
    
     \begin{align}
     \vdots & \notag\\
        \underline{\text{for }l_{n_0-1}^{T_r}}\text{: } & \notag \\
        \nabla l_{n_0}^{T_r} &= \frac{\partial T_{rp}}{\partial l_{n_0}^{T_r}} + \frac{\partial C_E}{\partial l_{n_0}^{T_r}}, \tag{32} 
    \end{align}
    
    \begin{multline*}
  \nabla l_{n_0}^{T_r} = \biggl[\frac{\partial T_{rp}}{\partial o_{n_T}^{T_r}}\frac{\partial o_{n_T}^{T_r}}{\partial f_{n_T}^{T_r}}\frac{\partial f_{n_T}^{T_r}}{\partial a_{n_T}^{T_r}}\frac{\partial a_{n_T}^{T_r}}{\partial o_{n_T-1}^{T_r}} \notag\\
  \frac{\partial o_{n_T-1}^{T_r}}{\partial f_{n_T-1}^{T_r}}\frac{\partial f_{n_T-1}^{T_r}}{\partial a_{n_T-1}^{T_r}}\frac{\partial a_{n_T-1}^{T_r}}{\partial o_{n_T-2}^{T_r}} \hdots  \notag\\
  \frac{\partial o_{n_1}^{T_r}}{\partial f_{n_1}^{T_r}}\frac{\partial f_{n_1}^{T_r}}{\partial a_{n_1}^{T_r}}\frac{\partial a_{n_1}^{T_r}}{\partial o_{n_0}^{T_r}}\frac{\partial o_{n_0}^{T_r}}{\partial f_{n_0}^{T_r}} \notag \frac{\partial f_{n_0}^{T_r}}{\partial a_{n_0}^{T_r}}\frac{\partial a_{n_0}^{T_r}}{\partial W_{n_0}^{T_r}} \biggr]\notag \\
  + \biggl[Q \frac{\partial f_{n_0}^{C}}{\partial o_{n_T}^{T_r}} \frac{\partial o_{n_T}^{T_r}}{\partial f_{n_T}^{T_r}}\frac{\partial f_{n_T}^{T_r}}{\partial a_{n_T}^{T_r}}\frac{\partial a_{n_T}^{T_r}}{\partial o_{n_T-1}^{T_r}} \notag\\
  \frac{\partial o_{n_T-1}^{T_r}}{\partial f_{n_T-1}^{T_r}}\frac{\partial f_{n_T-1}^{T_r}}{\partial a_{n_T-1}^{T_r}}\frac{\partial a_{n_T-1}^{T_r}}{\partial o_{n_T-2}^{T_r}} \hdots \notag\\
  \frac{\partial o_{n_1}^{T_r}}{\partial f_{n_1}^{T_r}}\frac{\partial f_{n_1}^{T_r}}{\partial a_{n_1}^{T_r}}\frac{\partial a_{n_1}^{T_r}}{\partial o_{n_0}^{T_r}}\frac{\partial o_{n_0}^{T_r}}{\partial f_{n_0}^{T_r}} \notag \frac{\partial f_{n_0}^{T_r}}{\partial a_{n_0}^{T_r}}\frac{\partial a_{n_0}^{T_r}}{\partial W_{n_0}^{T_r}} \biggr].\tag{33}
    \end{multline*}
    
    \item \textbf{Prior branch:}
    
     \begin{align}
    \underline{\text{for }l_{n_P}^{P}}\text{: } &  \notag\\
    \nabla l_{n_P}^{P} &= \frac{\partial C_E}{\partial l_{n_P}^{P}} \tag{34} \\
    &= \biggl[Q \frac{\partial f_{n_0}^{C}}{\partial o_{n_P}^{P}} \frac{\partial o_{n_P}^{P}}{\partial f_{n_P}^{P}}\frac{\partial f_{n_P}^{P}}{\partial a_{n_P}^{P}}\frac{\partial a_{n_P}^{P}}{\partial W_{n_P}^{P}} \biggr] \tag{35}
    \end{align}
    
    \begin{multline*}
    \underline{\text{for }l_{n_P-1}^{P}}\text{: } \\
    \nabla l_{n_P-1}^{P} = \biggl[Q \frac{\partial f_{n_0}^{C}}{\partial o_{n_P}^{P}} \frac{\partial o_{n_P}^{P}}{\partial f_{n_P}^{P}}\frac{\partial f_{n_P}^{P}}{\partial a_{n_P}^{P}}\frac{\partial a_{n_P}^{P}}{\partial o_{n_P-1}^{P}} \notag\\
  \frac{\partial o_{n_P-1}^{P}}{\partial f_{n_P-1}^{P}}\frac{\partial f_{n_P-1}^{P}}{\partial a_{n_P-1}^{P}}\frac{\partial a_{n_P-1}^{P}}{\partial W_{n_P-1}^{P}}\biggr] \tag{36}
    \end{multline*}
    
     \vspace{-5mm} 
    \begin{align}
        \vdots \notag
    \end{align}
    \vspace{-13mm} 
    
  \begin{multline*}
    \underline{\text{for }l_{n_0}^{P}}\text{: } \\
    \nabla l_{n_0}^{P} = \biggl[Q \frac{\partial f_{n_0}^{C}}{\partial o_{n_P}^{P}} \frac{\partial o_{n_P}^{P}}{\partial f_{n_P}^{P}}\frac{\partial f_{n_P}^{P}}{\partial a_{n_P}^{P}}\frac{\partial a_{n_P}^{P}}{\partial o_{n_P-1}^{P}} \notag\\
  \frac{\partial o_{n_P-1}^{P}}{\partial f_{n_P-1}^{P}}\frac{\partial f_{n_P-1}^{P}}{\partial a_{n_P-1}^{P}}\frac{\partial a_{n_P-1}^{P}}{\partial o_{n_P-2}^{P}} \hdots \\
   \frac{\partial o_{n_1}^{P}}{\partial f_{n_1}^{P}}\frac{\partial f_{n_1}^{P}}{\partial a_{n_1}^{P}}\frac{\partial a_{n_1}^{P}}{\partial o_{n_0}^{P}}\frac{\partial o_{n_0}^{P}}{\partial f_{n_0}^{P}} \notag \frac{\partial f_{n_0}^{P}}{\partial a_{n_0}^{P}}\frac{\partial a_{n_0}^{P}}{\partial W_{n_0}^{P}} \biggr] \tag{37}
    \end{multline*}  
    
\end{enumerate}

The gradient backpropagated by the Biceph module is the combination of gradients backpropagated by the triplet and prior branches, and can be calculated as follows:

\begin{align}
    \nabla{biceph} &= \nabla{l_{n_0}^{T_r}} + \nabla{l_{n_0}^{P}}. \tag{38}
\end{align}

Therefore, $\nabla{biceph}$ = 
\begin{multline*}
    \biggl[\frac{\partial T_{rp}}{\partial o_{n_T}^{T_r}}\frac{\partial o_{n_T}^{T_r}}{\partial f_{n_T}^{T_r}}\frac{\partial f_{n_T}^{T_r}}{\partial a_{n_T}^{T_r}}\frac{\partial a_{n_T}^{T_r}}{\partial o_{n_T-1}^{T_r}} \notag\\
  \frac{\partial o_{n_T-1}^{T_r}}{\partial f_{n_T-1}^{T_r}}\frac{\partial f_{n_T-1}^{T_r}}{\partial a_{n_T-1}^{T_r}}\frac{\partial a_{n_T-1}^{T_r}}{\partial o_{n_T-2}^{T_r}} \hdots  \notag\\
  \frac{\partial o_{n_1}^{T_r}}{\partial f_{n_1}^{T_r}}\frac{\partial f_{n_1}^{T_r}}{\partial a_{n_1}^{T_r}}\frac{\partial a_{n_1}^{T_r}}{\partial o_{n_0}^{T_r}}\frac{\partial o_{n_0}^{T_r}}{\partial f_{n_0}^{T_r}} \notag \frac{\partial f_{n_0}^{T_r}}{\partial a_{n_0}^{T_r}}\frac{\partial a_{n_0}^{T_r}}{\partial W_{n_0}^{T_r}} \biggr]\notag \\
  + \biggl[Q \frac{\partial f_{n_0}^{C}}{\partial o_{n_T}^{T_r}} \frac{\partial o_{n_T}^{T_r}}{\partial f_{n_T}^{T_r}}\frac{\partial f_{n_T}^{T_r}}{\partial a_{n_T}^{T_r}}\frac{\partial a_{n_T}^{T_r}}{\partial o_{n_T-1}^{T_r}} \notag\\
  \frac{\partial o_{n_T-1}^{T_r}}{\partial f_{n_T-1}^{T_r}}\frac{\partial f_{n_T-1}^{T_r}}{\partial a_{n_T-1}^{T_r}}\frac{\partial a_{n_T-1}^{T_r}}{\partial o_{n_T-2}^{T_r}} \hdots \notag\\
  \frac{\partial o_{n_1}^{T_r}}{\partial f_{n_1}^{T_r}}\frac{\partial f_{n_1}^{T_r}}{\partial a_{n_1}^{T_r}}\frac{\partial a_{n_1}^{T_r}}{\partial o_{n_0}^{T_r}}\frac{\partial o_{n_0}^{T_r}}{\partial f_{n_0}^{T_r}} \notag \frac{\partial f_{n_0}^{T_r}}{\partial a_{n_0}^{T_r}}\frac{\partial a_{n_0}^{T_r}}{\partial W_{n_0}^{T_r}} \biggr] \notag \\
  + \biggl[Q \frac{\partial f_{n_0}^{C}}{\partial o_{n_P}^{P}} \frac{\partial o_{n_P}^{P}}{\partial f_{n_P}^{P}}\frac{\partial f_{n_P}^{P}}{\partial a_{n_P}^{P}}\frac{\partial a_{n_P}^{P}}{\partial o_{n_P-1}^{P}} \notag\\
  \frac{\partial o_{n_P-1}^{P}}{\partial f_{n_P-1}^{P}}\frac{\partial f_{n_P-1}^{P}}{\partial a_{n_P-1}^{P}}\frac{\partial a_{n_P-1}^{P}}{\partial o_{n_P-2}^{P}} \hdots \\
   \frac{\partial o_{n_1}^{P}}{\partial f_{n_1}^{P}}\frac{\partial f_{n_1}^{P}}{\partial a_{n_1}^{P}}\frac{\partial a_{n_1}^{P}}{\partial o_{n_0}^{P}}\frac{\partial o_{n_0}^{P}}{\partial f_{n_0}^{P}} \notag \frac{\partial f_{n_0}^{P}}{\partial a_{n_0}^{P}}\frac{\partial a_{n_0}^{P}}{\partial W_{n_0}^{P}} \biggr] \tag{39}
\end{multline*}

\begin{multline*}
  =  \biggl[\frac{\partial T_{rp}}{\partial o_{n_T}^{T_r}}\frac{\partial o_{n_T}^{T_r}}{\partial f_{n_T}^{T_r}}\frac{\partial f_{n_T}^{T_r}}{\partial a_{n_T}^{T_r}}\frac{\partial a_{n_T}^{T_r}}{\partial o_{n_T-1}^{T_r}} \notag\\
  \frac{\partial o_{n_T-1}^{T_r}}{\partial f_{n_T-1}^{T_r}}\frac{\partial f_{n_T-1}^{T_r}}{\partial a_{n_T-1}^{T_r}}\frac{\partial a_{n_T-1}^{T_r}}{\partial o_{n_T-2}^{T_r}} \hdots  \notag\\
  \frac{\partial o_{n_1}^{T_r}}{\partial f_{n_1}^{T_r}}\frac{\partial f_{n_1}^{T_r}}{\partial a_{n_1}^{T_r}}\frac{\partial a_{n_1}^{T_r}}{\partial o_{n_0}^{T_r}}\frac{\partial o_{n_0}^{T_r}}{\partial f_{n_0}^{T_r}} \notag \frac{\partial f_{n_0}^{T_r}}{\partial a_{n_0}^{T_r}}\frac{\partial a_{n_0}^{T_r}}{\partial W_{n_0}^{T_r}} \biggr]\notag \\
  + \biggl\{\biggl[Q \frac{\partial f_{n_0}^{C}}{\partial o_{n_T}^{T_r}} \frac{\partial o_{n_T}^{T_r}}{\partial f_{n_T}^{T_r}}\frac{\partial f_{n_T}^{T_r}}{\partial a_{n_T}^{T_r}}\frac{\partial a_{n_T}^{T_r}}{\partial o_{n_T-1}^{T_r}} \notag\\
  \frac{\partial o_{n_T-1}^{T_r}}{\partial f_{n_T-1}^{T_r}}\frac{\partial f_{n_T-1}^{T_r}}{\partial a_{n_T-1}^{T_r}}\frac{\partial a_{n_T-1}^{T_r}}{\partial o_{n_T-2}^{T_r}} \hdots \notag\\
  \frac{\partial o_{n_1}^{T_r}}{\partial f_{n_1}^{T_r}}\frac{\partial f_{n_1}^{T_r}}{\partial a_{n_1}^{T_r}}\frac{\partial a_{n_1}^{T_r}}{\partial o_{n_0}^{T_r}}\frac{\partial o_{n_0}^{T_r}}{\partial f_{n_0}^{T_r}} \notag \frac{\partial f_{n_0}^{T_r}}{\partial a_{n_0}^{T_r}}\frac{\partial a_{n_0}^{T_r}}{\partial W_{n_0}^{T_r}}  \notag \\
  + Q \frac{\partial f_{n_0}^{C}}{\partial o_{n_P}^{P}} \frac{\partial o_{n_P}^{P}}{\partial f_{n_P}^{P}}\frac{\partial f_{n_P}^{P}}{\partial a_{n_P}^{P}}\frac{\partial a_{n_P}^{P}}{\partial o_{n_P-1}^{P}} \notag\\
  \frac{\partial o_{n_P-1}^{P}}{\partial f_{n_P-1}^{P}}\frac{\partial f_{n_P-1}^{P}}{\partial a_{n_P-1}^{P}}\frac{\partial a_{n_P-1}^{P}}{\partial o_{n_P-2}^{P}} \hdots \\
   \frac{\partial o_{n_1}^{P}}{\partial f_{n_1}^{P}}\frac{\partial f_{n_1}^{P}}{\partial a_{n_1}^{P}}\frac{\partial a_{n_1}^{P}}{\partial o_{n_0}^{P}}\frac{\partial o_{n_0}^{P}}{\partial f_{n_0}^{P}} \notag \frac{\partial f_{n_0}^{P}}{\partial a_{n_0}^{P}}\frac{\partial a_{n_0}^{P}}{\partial W_{n_0}^{P}} \biggr]\biggr\}, \label{appendix_eq40} \tag{40}
\end{multline*}

The value of $\nabla{biceph}$ can be viewed as the combination two different terms as shown in equation (\ref{appendix_eq40}). The  first term in the equation denotes the inter-slice information that the Biceph module bakcpropagates. Whereas, the second term represents the intra-slice information bakcpropagated by the Biceph module. The inter-slice information helps in performing better subject-wise classification whereas, the intra-slice information helps in performing better slice-wise classification.

\begin{algorithm*}
\small
\caption{Online triplet mining and loss calculation for Biceph-Net}\label{biceph_algo}
\textbf{Variables required:} \\
$H$ - number of subjects.\\
$m$ - number of 2D-MRI slices of each subject.\\
$W$ - number of subjects chosen in each batch.\\
$Z$ - number of slices of each $w \, \in \, W$.\\
$B$ - size of each batch = $W.Z$.\\
$F_{T_r}$ - feature vector from triplet branch.\\
$F_C$ - feature vector from concatenate branch.\\
$M_{T_r}$ - margin value for the triplet loss.\\
$epochs$ - total number of epochs to run.\\
$d(a,b)$ - function to calculate euclidean distance between vectors $a$ and $b$.\\
$C$ - $C=1$  for binary classification problems (CN vs AD or MCI vs AD) and $C=3$ for multiclass classification problem (CN vs MCI vs AD).\\
\textbf{Algorithm:}\\
\For{$j \, \in epochs$}{
Select $W$ subjects uniformly at random.\\
Select $Z$ slices for each $w \, \in \, W$ without replacement.\\
Perform forward pass for a batch of size $B$.\\
Let $O_{T_r}$ be an output matrix  of dimension $B\times F_{T_r}$ for the forward pass from the triplet branch.\\
Let $O_C$ be an output matrix of dimension $B \times C$ from the concatenate branch.\\
From  $O_{T_r}$, compute a distance matrix $D$ of dimension $B \times B$.\\
\For{each row $r$ in $O_{T_r}$}{
Find $e$, $i$ such that $d(r,e) \, < \, d(r, i) \, <d(r, e) + M_{T_r}$. Here, $r$ is the anchor, $e$ is the positive sample and $i$ is the semi-hard negative sample.\label{step1}\\

Calculate triplet loss for the entire batch $B$ using semi-hard triplets as found in step \ref{step1}.\\
}

For a batch $B$, compute the CE loss.\\
}
  
\end{algorithm*}

\section{Models for comparison}
In this section we describe about the models chosen for comparison with the proposed methods. We focus on the models that can utilize 2D-MRI scans for AD classification and do not consider 3D-CNN models. It is important to note that a deep neural network has many different hyperparameters that affect its performance. In order to maintain fairness in comparison the best hyperparameters after performing a manual hyperparameter tuning on the validation set were chosen for all the models used for comparison. The two categories of models we choose for comparison are as follows: 
\begin{enumerate}
\item \textbf{2D-CNN models:} 

    \begin{enumerate}
     \item \textbf{Eight-layer CNN:} We follow the exactly similar implementation as done in \citeA{wang2018classification}. 
     
        \item \textbf{Triplet network:} The flattened features from the backbone are passed onto a 64 dimensional dense layer, which then is connected to an L2-normalization layer. The features from L2-normalization layer are passed onto the Triplet-loss layer \citeA{hermans2017defense}. We term this as VGG-Triplet. We perform final classification using the $K$-Nearest Neighbour Algorithm with various different values of $K$. For slice-wise classification, we find $K$ closest neighbours using euclidean distance. For subject wise classification, we first get the predicted labels for each image and them using majority voting, i.e., if 43 or more slices (in case of axial and sagittal) and 56 or more slices (in case of coronal) of a particular subject has a certain label then that label is the predicted label for that subject. The value of $K$ is found using crossvalidation.
       
    \end{enumerate}
    
\item \textbf{CNN-RNN models:} In order to maintain fairness in comparison, the backbone architecture is chosen as mentioned in the main manuscript figure 5. 

    \begin{enumerate}
        \item \textbf{LSTM:} \label{comp:lstm} The flattened features from the backbone are passed onto an LSTM layer with 50 nodes \citeA{dua2020cnn}. The features from the LSTM layer are passed onto a dense layer with 10 nodes, which is, connected to the final classification layer with 1 node and sigmoid activation function (for binary classification) and 3 nodes with softmax activation function (for multiclass classification). We term this as VGG-LSTM.
        \item \textbf{Stacked bidirectional LSTM:} \label{comp:sbi-lstm} The flattened features from the backbone are passed onto two bidirectional LSTM layers (sBi-LSTM) \citeA{el2020alzheimer}, each having 50 nodes. The features from sBi-LSTM are passed onto final classification layer which is similar as in the LSTM case \ref{comp:lstm}. We term this as VGG-sBiLSTM. 
        \item \textbf{Conv-LSTM:}\label{comp:conv-lstm} The features from the backbone architecture are passed onto two Conv-LSTM \citeA{shi2015convolutional} layers, each having 32 filters with 3$\times$3 kernel. The features from the final Conv-LSTM layer are flattened and passed onto two dense layers similar to the LSTM scenario \ref{comp:lstm}.
    \end{enumerate}
    
\end{enumerate}

Further details about hyperparameter configurations for comparative models can be found in table \ref{model_config} of the Appendix material.

\begin{figure*}
\centering     
\subfigure[]{\label{fig:pca1_tr_axial_cn_vs_ad}\includegraphics[width=5cm]{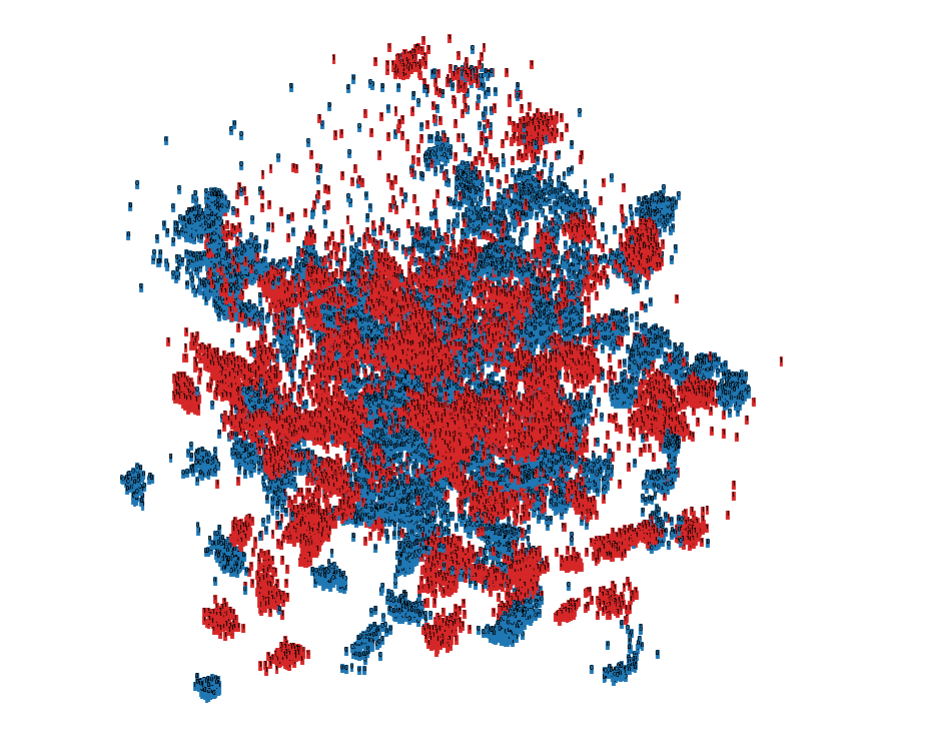}}
\subfigure[]{\label{fig:pca1_tr_axial_mci_vs_ad}\includegraphics[width=5cm]{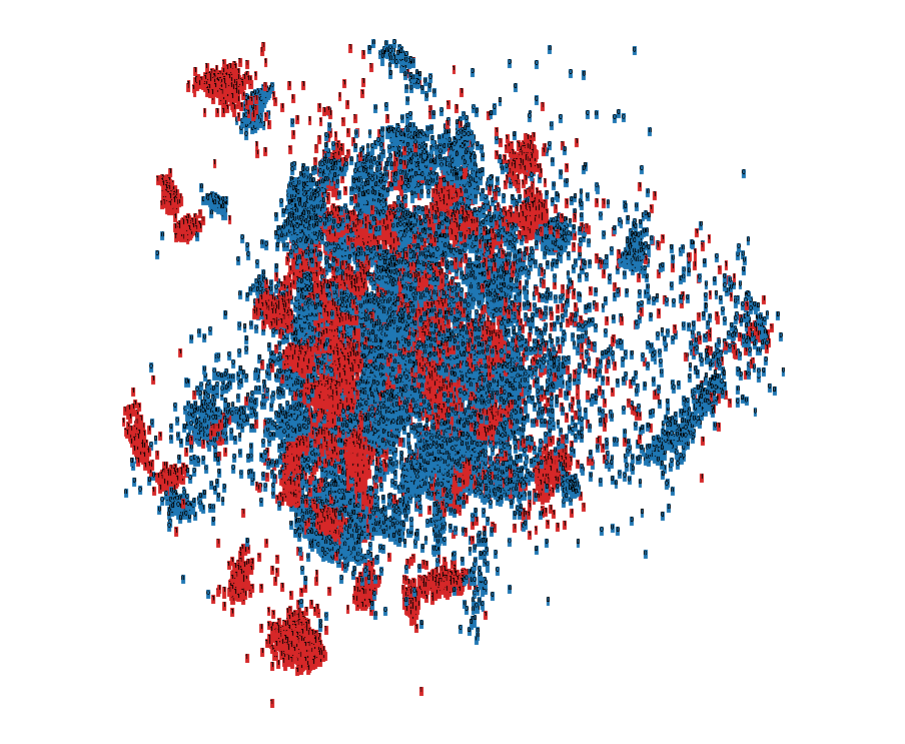}}
\subfigure[]{\label{fig:pca1_tr_axial_multiclass}\includegraphics[width=5cm]{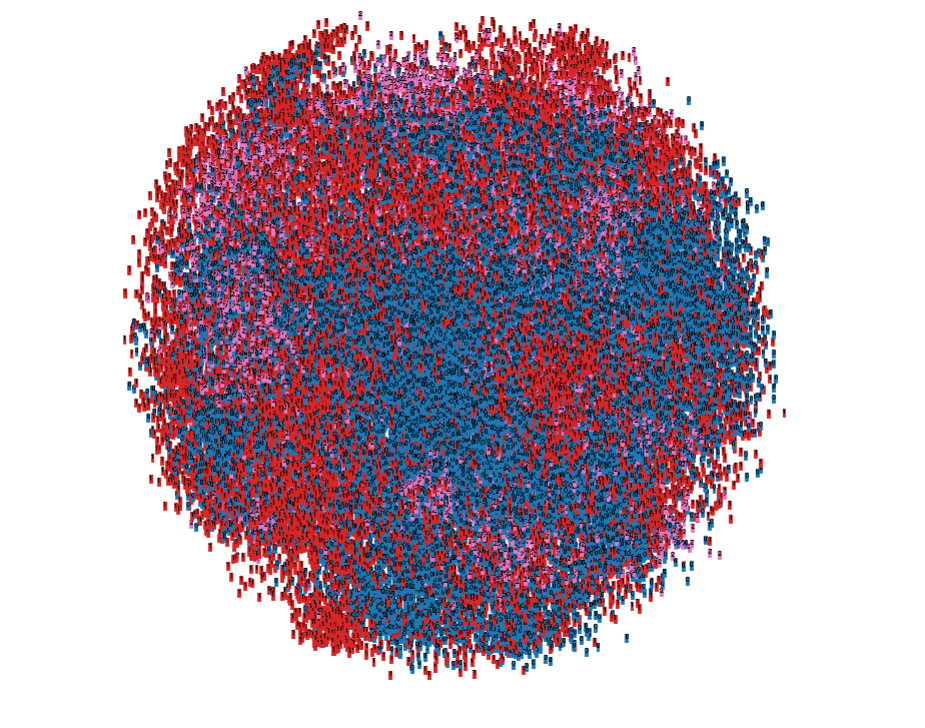}}
\subfigure[]{\label{fig:pca1_bi_axial_cn_vs_ad}\includegraphics[width=5cm]{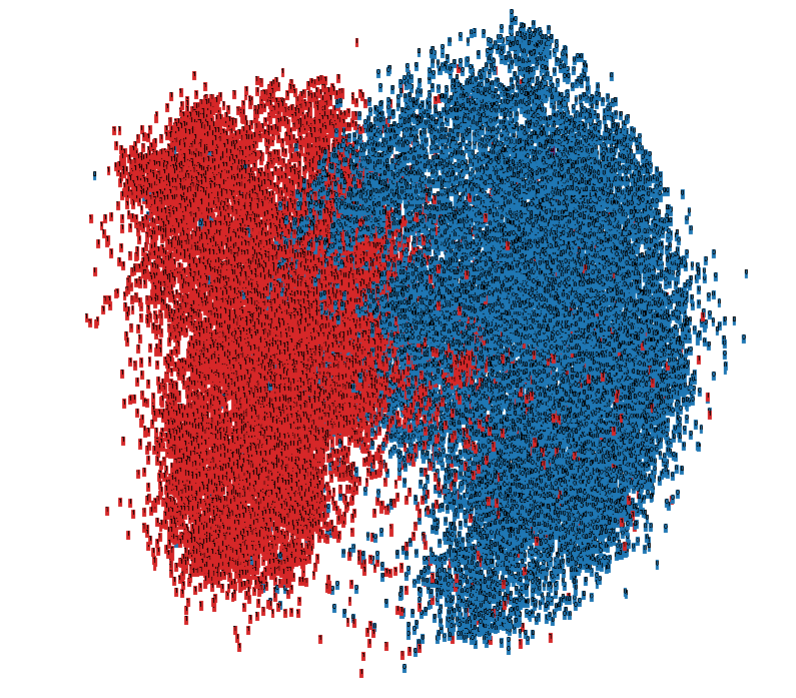}}
\subfigure[]{\label{fig:pca1_bi_axial_mci_vs_ad}\includegraphics[width=5cm]{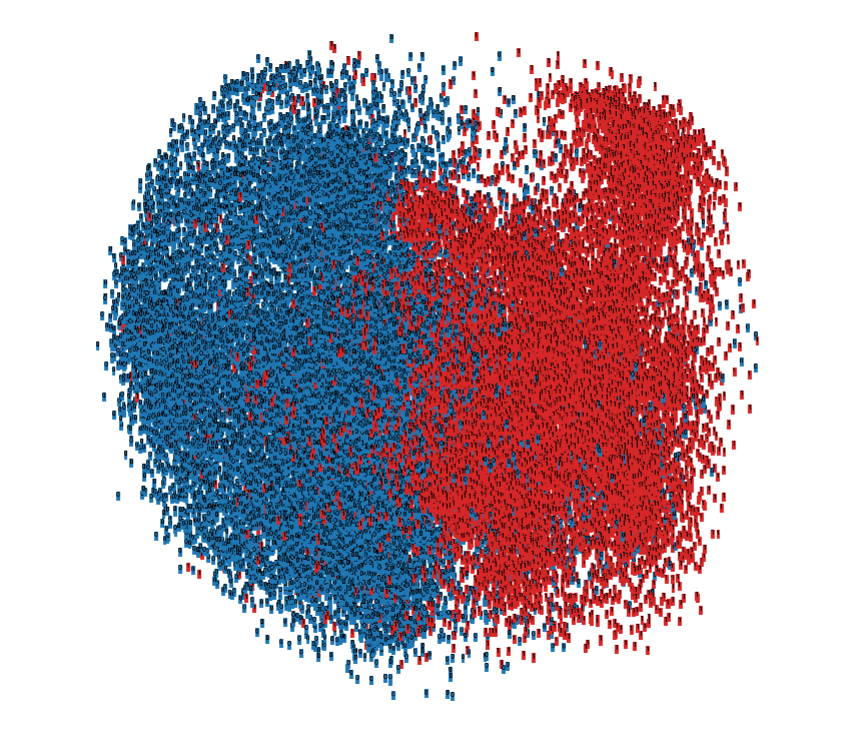}}
\subfigure[]{\label{fig:pca1_bi_axial_multiclass}\includegraphics[width=5cm]{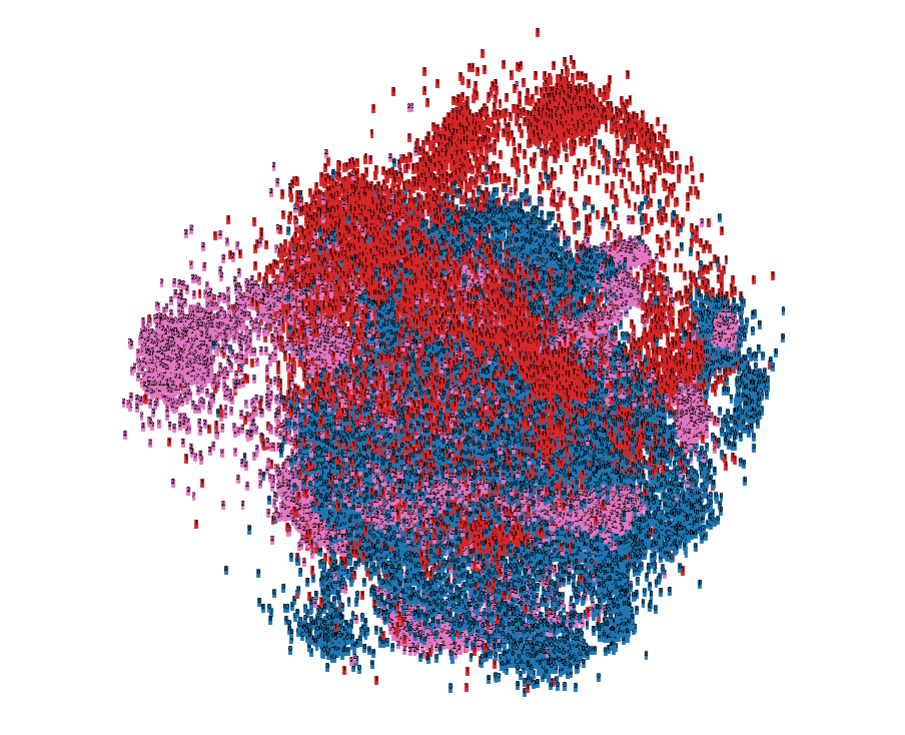}}
\caption{PCA Visualization of feature embeddings of Triplet network and Biceph network. Figures (a)--(c) represent embeddings for VGG-Triplet network on CN vs AD, MCI vs AD and CN vs MCI vs AD respectively. Figures (d)--(f) represent embeddings for Biceph-Net for CN vs AD, MCI vs AD and CN vs MCI vs AD respectively.}
\label{fig:pca}
\end{figure*}

\begin{figure*}
\centering     
\subfigure[]{\label{fig:tsne_tr_axial_cn_vs_ad}\includegraphics[width=5cm]{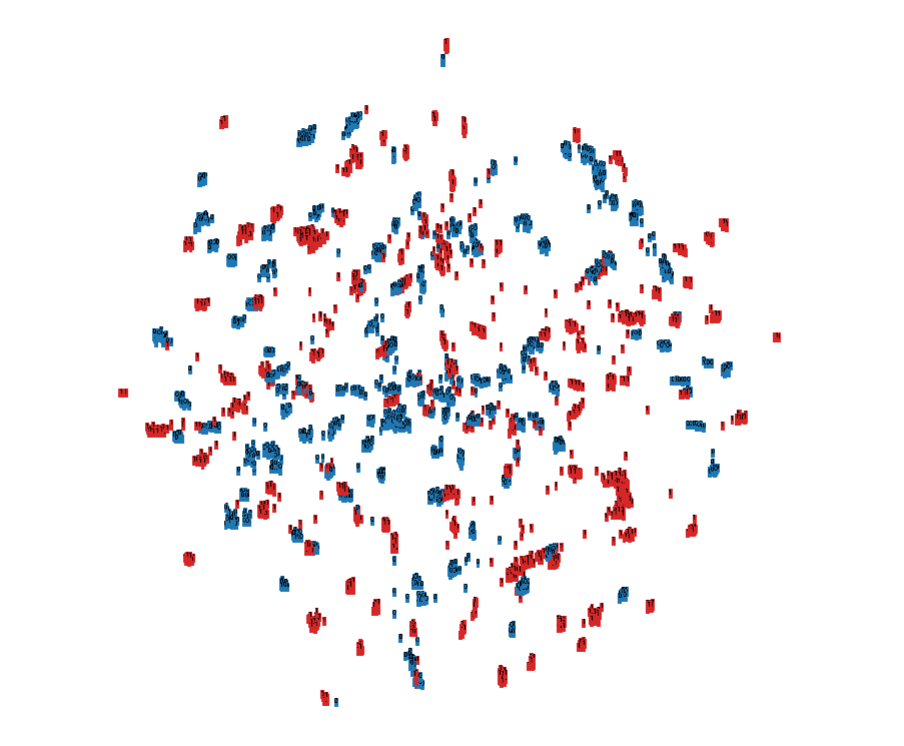}}
\subfigure[]{\label{fig:tsne_tr_axial_mci_vs_ad}\includegraphics[width=5cm]{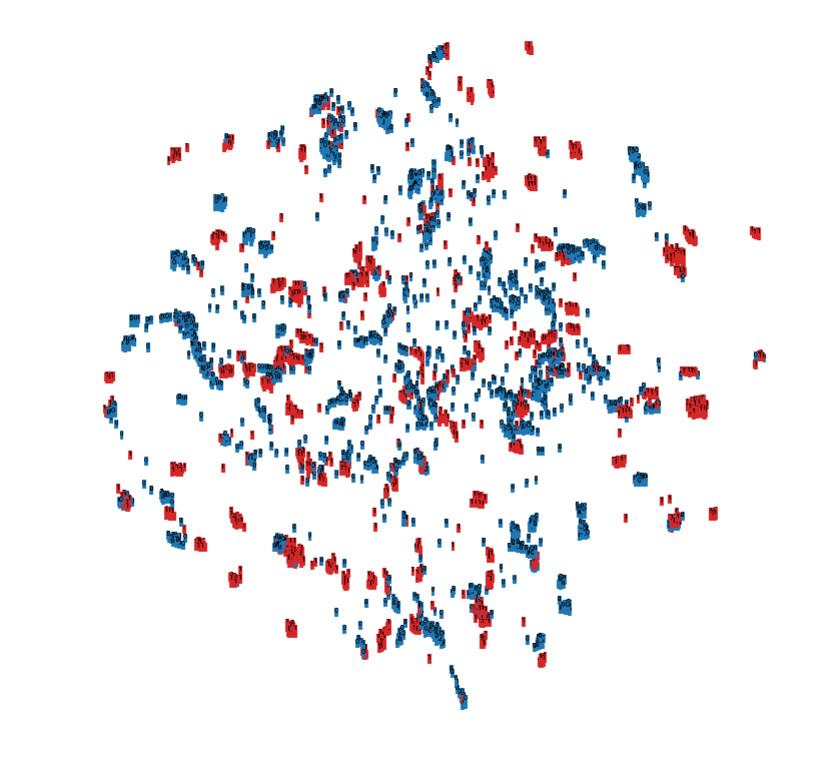}}
\subfigure[]{\label{fig:tsne_tr_axial_multiclass}\includegraphics[width=5cm]{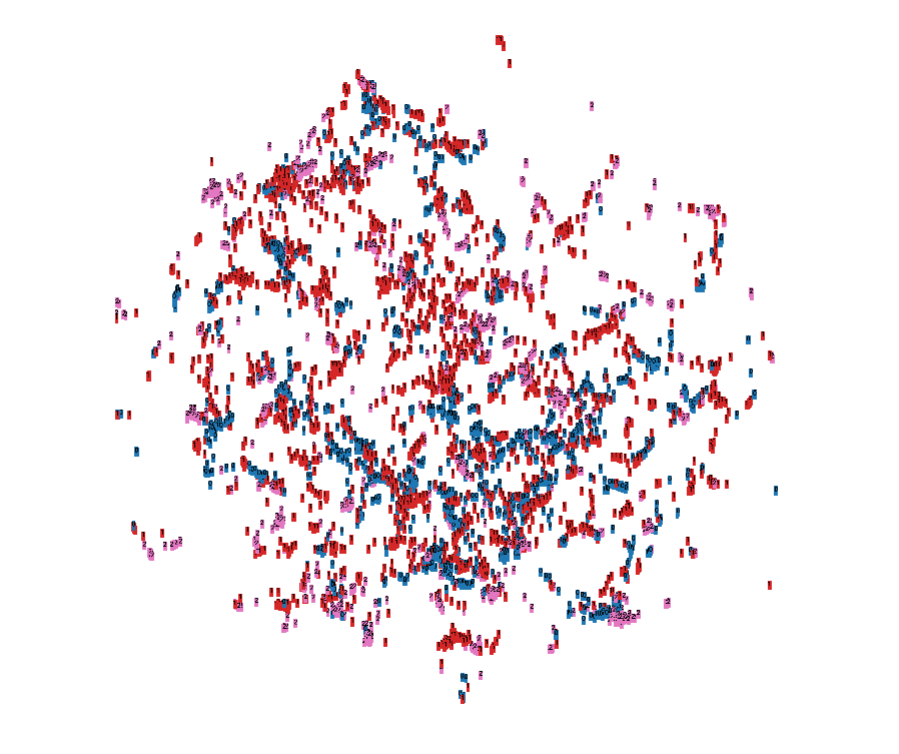}}
\subfigure[]{\label{fig:tsne_bi_axial_cn_vs_ad}\includegraphics[width=5cm]{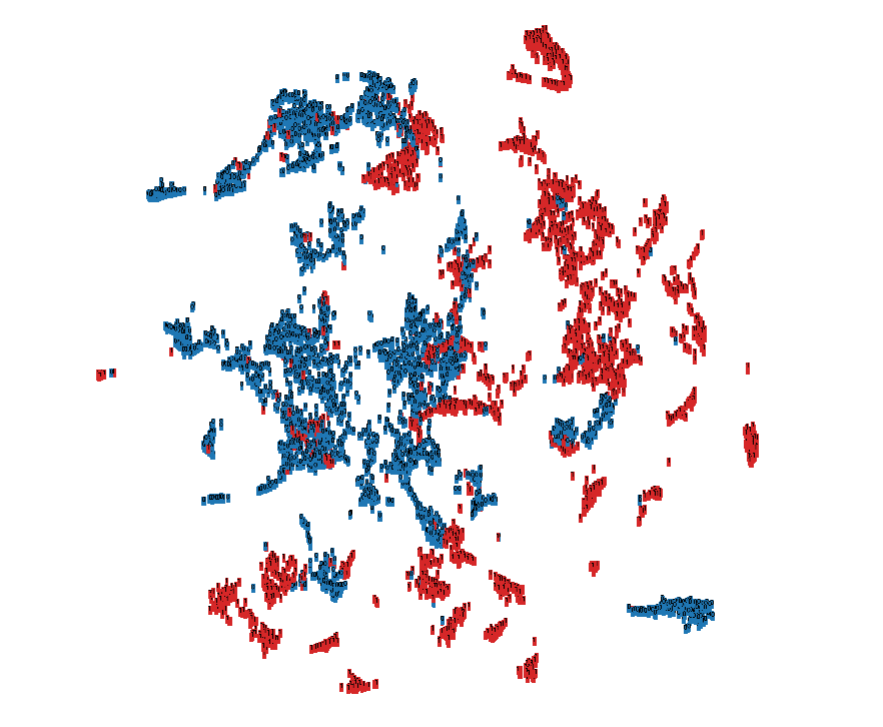}}
\subfigure[]{\label{fig:tsne_bi_axial_mci_vs_ad}\includegraphics[width=5cm]{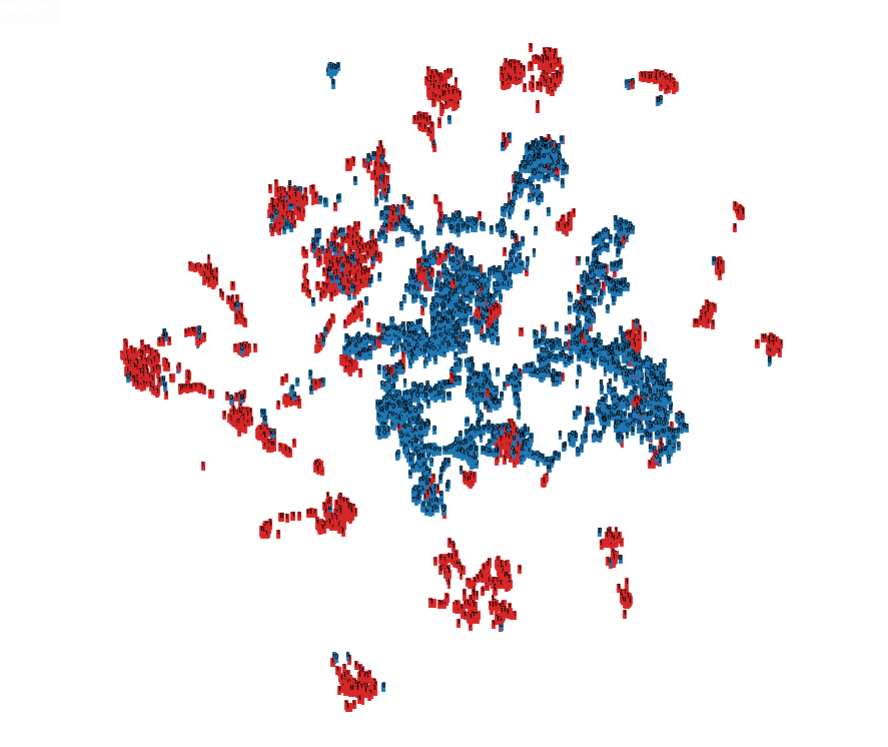}}
\subfigure[]{\label{fig:tsne_bi_axial_multiclass}\includegraphics[width=5cm]{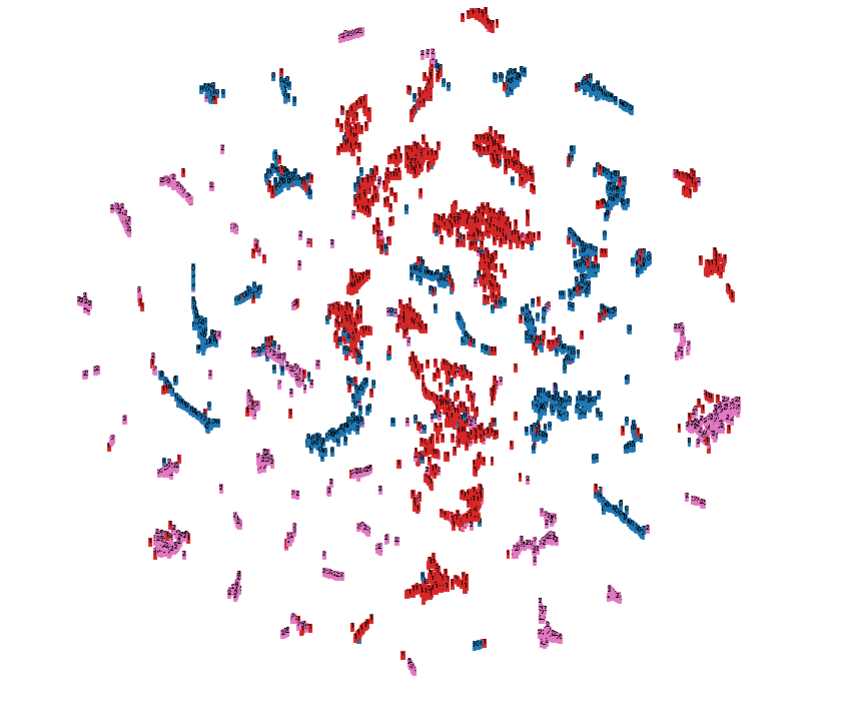}}
\caption{T-sne Visualization of feature embeddings of Triplet network and Biceph network. Figures (a)--(c) represent embeddings for VGG-Triplet network on CN vs AD, MCI vs AD and CN vs MCI vs AD respectively. Figures (d)--(f) represent embeddings for Biceph-Net for CN vs AD, MCI vs AD and CN vs MCI vs AD respectively.}
\label{tsne_biceph}
\end{figure*}

\end{appendices}


\bibliographystyleA{IEEEtran}
\bibliographyA{bibliographysupp}

\end{document}